\documentclass[%
 reprint,
 amsmath,amssymb,
 aps,
superscriptaddress,
]{revtex4-1}

\usepackage{longtable}
\usepackage{graphicx}
\usepackage{dcolumn}
\usepackage{bm}

\usepackage{enumitem}
\usepackage{fixltx2e}
\usepackage{hyperref}
\usepackage{listings}
\usepackage{color}
\usepackage[caption=false]{subfig}

\begin{document}

\title{Characterizing the structural diversity of complex networks across domains} 

\author{Kansuke Ikehara}
\email{kansuke.ikehara@colorado.edu}
\affiliation{Department of Computer Science, University of Colorado, Boulder, CO, USA}

\author{Aaron Clauset}
\email{aaron.clauset@colorado.edu}
\affiliation{Department of Computer Science, University of Colorado, Boulder, CO, USA}
\affiliation{BioFrontiers Institute, University of Colorado, Boulder, CO, USA}
\affiliation{Santa Fe Institute, Santa Fe, NM, USA}

\begin{abstract}
The structure of complex networks has been of interest in many scientific and engineering disciplines over the decades. A number of studies in the field have been focused on finding the common properties among different kinds of networks such as heavy-tail degree distribution, small-worldness and modular structure and they have tried to establish a theory of structural universality in complex networks. However, there is no comprehensive study of network structure across a diverse set of domains in order to explain the structural diversity we observe in the real-world networks. In this paper, we study 986 real-world networks of diverse domains ranging from ecological food webs to online social networks along with 575 networks generated from four popular network models. Our study utilizes a number of  machine learning techniques such as random forest and confusion matrix in order to show the relationships among network domains in terms of network structure. Our results indicate that there are some partitions of network categories in which networks are hard to distinguish based purely on network structure. We have found that these partitions of network categories tend to have similar underlying functions, constraints and/or generative mechanisms of networks even though networks in the same partition have different origins, e.g., biological processes, results of engineering by human being, etc. This suggests that the origin of a network, whether it's biological, technological or social, may not necessarily be a decisive factor of the formation of similar network structure. Our findings shed light on the possible direction along which we could uncover the hidden principles for the structural diversity of complex networks.
\footnote{This document was first published as K. Ikehara, ``The Structure of Complex Networks across Domains.'' MS Thesis, University of Colorado Boulder (2016).}
\end{abstract}

\maketitle

\section*{Introduction} 
Almost every scientific and engineering discipline deals with data that comes from some sort of experimental observations. Traditionally, such data are expressed as numbers, which may represent temperature, velocity, or voltage, and methodologies that analyze the data have been established for over hundreds of years. A \textit{graph} or \textit{network}, which is a kind of data representation describing the relations among some entities such as persons, molecules, animals, logical gates, etc., has been used as a new way to approach, interpret and solve real-world problems in the last decades. Social science, for instance, has witnessed the power of network analysis with the recent emergence of online social network services such as Facebook, Twitter, LinkedIn, etc.\cite{Kleinberg:1}. Previously invisible social phenomena at the scale of off-line social networks, which usually consist of tens to hundreds of nodes at most, have been observed in large scale online social networks with the abundance of online data, faster and more efficient computational resources along with advancements of graph algorithms.

	Biological sciences use networks as a tool to dissect biological, chemical and ecological processes in order to gain insight into the functionality of such processes. These include brain that consists of networks of neurons \cite{BrainNetwork}, complex metabolic reactions within a human's cells and relationships between malfunctioning metabolic processes and human diseases \cite{MetabolicNetworkAndDiseases} and the effect in biodiversity, that might result from perturbation in an ecological network, such as food-web, mutualistic network, etc \cite{EcologicalNetwork}. Engineering systems such as the Internet \cite{Internet}, power grids \cite{PowerGrid}, water distribution networks \cite{WaterDistribution}, transportation networks \cite{Train}, etc. have also been investigated using network analysis tools for constructing more efficient and robust systems. 
	
	\subsection*{Common patterns across domains}
	The term \textit{complex network} depicts an essential difference from ordinary graphs having some kind of regular structure that have been studied in the field of mathematics for a long time: the structure of real-world networks almost always exhibits an unusual pattern that greatly deviates from the regular structure and this seemingly irregular and complex structure can often be a clue to the underlying mechanism of the process of interest.  For example, the unusual density of triangles in social networks implies an underlying mechanism in our society: the formation of our social circles tends to be made by local interactions, such as introducing your friend to another friend in your circle that results in a new connection between your friends, making a triangle in the circle. There have been a number of studies that have investigated the structure of complex networks of diverse fields and connections between the structure and the underlying mechanism of the process \cite{Strogatz2001, Newman03thestructure, StatisticalMechanics, boccaletti06}. In the following sections we discuss the theories in the structure of complex networks in great detail.
	
	The structure of a network can be characterized in a number of ways, but there are three structural properties that are found to be common in many types of networks: the skewed degree distribution, small-worldness and community structure.
	
	\medskip \paragraph*{Degree distributions.} The degree of a node is a measure of how many edges are connected to a particular node, and the probability distribution $p(k)$ over all nodes essentially describes the ``unfairness'' in the network. If all of the nodes in a network have the exact same number of connections, the fairest case, then $p(k)$ behaves like Kronecker delta, where $p(k)$ has a value $1$ only at a specific degree $k$. Or if it is less fair, one may observe a narrow Gaussian-like distribution around an average degree. In the real world network, however, this almost never happens. What is observed most of the time instead, is a very skewed degree distribution in which most of the nodes have a few connections incident upon them, and very few nodes have a disproportionally large number of connections with them.
	
	The \textit{power-law} distribution is one candidate distribution for describing the observed phenomenon, and a network having this distribution is often called \textit{scale-free} network \cite{Barabasi99emergenceScaling}. A number of networks  from diverse fields, such the Internet, metabolic reactions, World-Wide-Web, etc.\ have been claimed to be a scale-free network. However, one needs to be careful in order to validate if a network of interest indeed has the power-law degree distribution and it appears that a number of such claims need more statistically valid treatment such as the one proposed by Clauset \textit{et al.} for justification \cite{Clauset:PowerLaw}. The disproportionality of the skewed degree distribution indicates the existence of  \textit{hubs}, \textit{cores} or \textit{elites} in a network. Such important nodes in a network can play a critical role for a network to function properly and the failure of such nodes may result in a catastrophe \cite{ScaleFreeAttack}.
	
	\medskip \paragraph*{Geodesic paths.} The next common structural characteristic of networks is \textit{small-worldness}, first conceptually introduced by a Milgram's experiment \cite{Milgram} and mathematically proposed in a paper by Watts and Strogatz \cite{watts1998cds}. Watts and Strogatz proposed a random network model that, depending on a parameter setting, produces a network which has the following properties: (i) the high density of triangles, implying that if three nodes are connected, it is likely that those three nodes actually compose a triangle; (ii) low average distance between a pair of nodes, which indicates that from any node it is just a few steps, in average, to reach any other node in the network. These properties are accomplished by the existence of ``long-range'' connections bridging together pairs of nodes topologically far away from each other.
	
	The small-worldness holds seemingly contradicting properties together: the large degree of local-ness exhibited as a large clustering coefficient value and the large degree of global-ness expressed as a small value of mean geodesic distance.  These properties make small-world networks a very efficient system for information flow \cite{SmallWorldEfficiency} and synchronizing coupled oscillators in the network \cite{SmallWorldSynchronization}. Note that the small-worldness itself is orthogonal to the skewed degree distribution, meaning that so-called small-world networks can be constructed without having the heavy-tail degree distribution. The definition of small world network presented above is, however, not generally used. Most of the researchers today regard small-world property as simply the low average pairwise distance between nodes that grows approximately as $O(\log n)$.

	\medskip \paragraph*{Community structure.} Many of the real-world networks contain \textit{modules} or \textit{communities} in which nodes are densely connected but among which there are sparse edges running. This is called \textit{community structure} and there has been a number of studies investigating communities and inventing a new algorithm for detecting communities in networks \cite{Modularity1, Modularity2, ModularityReview}. The communities in a network found by algorithms often correspond to the functional units of the network: a unit of chemical reactions producing vital chemical product in the metabolic network, a group of densely connected neurons taking charge of a cognitive function, such as language and visual processing, and a group of scientists working together in the same field.

	\subsection*{Stylized structural ``facts''}
The properties such as skewed-degree distribution, small-worldness and community structure are found in networks of various kinds, but they alone cannot explain the diversity of networks in terms of network structure. It has been believed by a number of researchers that some classes of networks have a set of distinguishing structural features that makes the specific network class ``stand out'' among others. Here, we show some examples of network class that have distinguishing structural features, including social networks, brain networks and subway networks.

\medskip \paragraph*{Social networks.}
Social networks, regardless of whether or not they are off-line as we see in our daily lives or online like Facebook, have long been known for having  distinguishing structural features: clustering and positive degree assortativity \cite{AssortativeMixing,WhySocialNetworks, Mislove:2007:OnlineSocial}. The large degree of clustering indicates the high probability of one's friends being friends to each other and the positive degree assortativity shows a tendency that high-degree nodes (nodes having many connections) connect to other high-degree nodes while low-degree nodes connect to other low-degree nodes. Real-world networks, \textit{except} social networks, in general exhibit low clustering and negative degree assortativity that are almost in accordance with ones of their randomized counterparts or null models having the same degree distribution.

Newman has argued~\cite{WhySocialNetworks} that negative degree assortativity is a natural state for most networks. Therefore, in order for a network to have positive degree assortativity, it needs a specific structure that favors the assortativity. The community structure of social networks, as mentioned in the paper, is the key aspect as to why they exhibit the properties, namely clustering and positive degree assortativity. People or nodes in social networks usually belong to some sort of groups or communities and people in the same community are likely to know each other. This community membership yields the high clustering in the social network.  For positive degree assortativity, the size of the communities may play an essential role: individuals in a small group can only have low degree whereas individuals in a large group can, potentially, have much larger degree. This approximate correspondence of degree in the groups is essentially degree assortativity itself.

\medskip \paragraph*{Brain networks.}
Brain networks, often referred to as \textit{connectome}, have been in the focus of neuroscience in the last decade in order to solve the long-standing scientific question: \textit{How does a brain work?}  The field that studies extensively brain networks using a tool set from mathematics, computer science, etc., is called \textit{connectomics} and one of the recent research topics in this field is investigating how a topology of a brain network affects the brain's function \cite{ComparativeConnectome}.

A number of studies in the topological features of brain networks show that: (i) they are highly modularly structured; (ii) they have topological connections that are anatomically long, thus come with the high cost of wiring, and yet make a brain network topologically small, like a small-world network; and (iii) they have a core of highly connected nodes, called \textit{rich-club} in some literatures, that connect modules across the network together. The modular structure in a brain network is found to be correlating with a discrete cognitive function of a brain such as processing visual signals from eyes and audio signals from ears. The existence of topologically short yet anatomically long connections enables a brain to have an efficient way to process the information flowing on the network. Furthermore the rich club of a brain network plays an important role of integrating information that is produced and processed in different parts of the brain and this integration of information enables animals including human beings to do complex tasks \cite{BrainEconomy,Crossley09072013}.

\medskip \paragraph*{Transportation networks.}
Networks of subways in the major cities around the world have an interesting common feature unique to them, that is \textit{core-branch} structure \cite{Train}. They have a ring-shape connections of stations and dense connections therein, referred to as the ``core'' of the network. In the core, stations are relatively densely connected to each other, enabling residents of a city to move around quickly. From the ring of the core, branches radiate outward connecting stations far from the center of the city. This structural feature is the result of balance between an efficiency of flow of people and cost for constructing rail lines between stations \cite{SpatioalNetwork1,SpatioalNetwork2}.

	\subsection*{Distinguishing different ``classes'' of networks}
	 In the last decades, researchers from the wide range of fields including biology, social science and physics, have been interested in if there is any structural difference between different classes of networks, if there is a set of structural features unique to a specific network class and if there is any ``family'' of networks in which networks of different classes share the same structural patterns. These questions are usually converted into a problem of comparing and classifying networks according to some criteria and these criteria span from a simple feature such as clustering coefficient, to more complicated ones such as network motifs, which are going to be explained later.
	
	One of the earliest works in network comparison and classification was done by Faust and Skvoretz \cite{Faust.Skvoretz2002Comparing}, in which they compared various kinds of offline social networks, such as grooming relationships among monkeys, co-sponsorship among U.S. Senate of 1973-1974 and so on. Their comparison of networks is based on a statistical model that incorporates parameters each describing a characteristic network structural feature, for example the frequency of a cyclic triangle (here the networks are \textit{directed}, meaning that edges have directionality). The statistical model essentially predicts the existence of an edge $(i,j)$ in a network based on the parameters and the authors used such a statistical model ``trained'' on a network for predicting an edge in a different network. Their assumption is that if two networks are similar, a model trained on one of the pair should predict well an edge existence in another network. With this assumption they define the Euclidian distance as a function of a summation over all edge predictions, construct a distance matrix and project it onto a two dimensional space using a technique called Correspondence Analysis, which is similar to Principal Component Analysis. They have found that what makes networks similar in terms of structure is the property of edges, namely a kind of relation. For example, networks describing agonistic relations, regardless of kinds of species, exhibit the similar network structure. Although this study was pioneering in graph comparison and classification, it only focused on offline social networks, which themselves are a very narrow field of study.
	
	The breakthrough in graph comparison and classification came along a series of papers by Milo \textit{et al.} that introduced the idea of \textit{network motifs} and ``super family'' of networks \cite{Milo_motif, Milo_SuperFamily}. Network motifs are essentially patterns of frequent sub-graph in a network compared to its randomized networks having the same degree distribution\cite{Milo_motif}. They have shown that each category of network, such as gene regulation (transcription), food webs, electronic circuits, etc., has distinct network motifs. In many cases the distinction of patterns indicates the functional difference in those networks. Furthermore, they revealed the existence of super families of networks that are groups of network categories having the highly convergent motif profiles. These studies are, as far as we know, the pioneers which investigated a diverse set of networks from different domains and found the underlying similarities among the network categories. Nevertheless, this study is far from proving to be a general theory as it only investigated 35 networks of a few categories. 
	
	One of the most recent studies in graph comparison and classification investigated 746 networks and constructed a taxonomy of networks \cite{Onnela_Taxonomy}.  Onnela \textit{et al.} used a technique called \textit{mesoscopic response functions} (MRFS) that essentially describes the change of a specific functional value related to the community structure of a network with respect to a parameter $\xi \in [0,1]$. Each network has its own MRFS and the authors calculated the distance between networks which is defined as an area of difference between two networks' MRFS. Their framework successfully identified groups of networks that are similar in terms of community structure. The drawback of study is, however, the fact that the metric they have used for clustering the networks, namely modularity, is implicitly correlated with the size of a network that is a very strong distinguishing feature for classification of networks.

	\subsection*{Quantifying the structural diversity of networks}
	As we have seen, a number of studies have been conducted in order to find groups or super families of networks that have similar structure. However, only few of them have investigated the fundamental concepts that create the structural differences of networks and none of them have done so with a comprehensive set of complex networks. With the abundance of available network data of various kinds and scales and the techniques from the field of machine learning, we could tackle a fundamental yet unexplored question:
	\begin{center}
	 \textit{What structural features distinguish different types of networks?}
	\end{center}
	 Or more generally:
	\begin{center}
	 \textit{What drives structural diversity among all complex networks?} 
	\end{center}
	
	 There have been a number of studies trying to discover or formulate an idea that explains the structural \textit{universality} of networks, including the skewed degree distribution, small-world networks, community structure and so on. There is, however, no general theory that explains the structural \textit{diversity} of complex networks across a number of domains/fields. The aim of this paper is to establish a theory that explains the structural diversity of complex networks. Below are the three questions we have formulated as research objectives:

\begin{enumerate}
	\item \textit{What aspects of network structure do make a specific category of network different from others?}
	
	This question is, in some extent, an extension to the studies investigating social networks' distinguishing structural features. For example, what kinds of network structure are distinguishing, say metabolic networks from the other kinds? As far as we know, few previous studies have done extensive investigation in finding distinctive characteristics of specific kinds of networks compared to other kinds except social networks.
	
	\item \textit{Are there any sets of network categories that are inherently indistinguishable from each other based on network structure?} 
	
	This question asks if there is any structural similarity between different kinds of networks. We, however, use the word ``indistinguishable'' in stead of ``similar'' since we try to observe the commonality from a confusion matrix of a classifier, where a misclassified instance is considered to be indistinguishable from a class it was classified as because it is so similar to other instances of the wrong class, the algorithm fails to label the instance correctly.
	
	\item \textit{If two networks of different categories are indistinguishable by network structure, are their mechanisms of underlying processes the same? And vice versa.}
	
	This question is essentially all about elucidating the meta-structure among the network domains. What if two very distinct domains of networks, say biological and technological ones, exhibit very similar network structure and a classifier misclassifies them many times? Is this because their underlying network generative processes, or their processes on networks themselves are the same? Answering this question helps us understand a mechanism for the formation of a specific network structure.
	
\end{enumerate}

\medskip
\paragraph*{Our contributions.} The contribution of answering those questions comes in two ways: first it gives us a general conceptual framework upon which networks are studied across domains. Previous studies have only looked at a single category or multiple categories as one category, ignoring the relationships between categories. By studying networks across domains, one could find general theories of networks or test a hypothesis across domains in a more plausible way.  For instance, one could test validity of a network model across all of domains and find out which domains the network model can explain well; second it gives us the knowledge base upon which various network-related algorithms could be constructed and tuned properly.

Many practical graph algorithms take no assumption in domain-specific network structure. It may be possible, however, to construct a new algorithm which runs faster and performs more efficiently on a specific kind of networks by taking into account of such domain-specific knowledge. For example, if there exists any unique network structural property of recommendation networks,  it may be applied to construct or fine-tune a recommendation engine that utilizes knowledge of unique network structure. As the size of some real-world networks has grown to an unprecedented scale, domain-specific knowledge in network structure may be a key to analyze such large networks in a faster and more efficient way.

In this paper, we study 986 of various kinds of real-world networks, ranging from ecological food-web to online social network to digital circuit along with 575 of synthetic networks generated from four different models. We extract network statistics as features of a network, construct a high-dimensional feature space, where each axis corresponds to one of the network features, and map each network onto the feature space. We then train a machine learning algorithm called \textit{Random Forest} with the training set of network data in order to learn the function that relates structural features of networks to class labels, namely network domains and sub-domains.

As the category distribution is skewed, meaning that some categories of networks have larger number of instances than the other minority categories, we try several sampling methods and show the effect of each methodology. We construct confusion matrices based upon classification results and proceed to analyze the misclassification with which we can answer research questions explained previously. We then conclude with discussion based on several hypotheses and experimental results along with some ideas for future work.

\section*{Formal Definitions}
	Formally, a network or graph is a mathematical object consisting of sets of edges (arcs) and nodes (vertices), which can be written as $G = (E,V)$. In many cases, an \textit{adjacency matrix} is used as a representation of a network, where each element of the matrix $A_{ij}$ takes a binary value, $1$ for presence of an edge between nodes $i$ and $j$, and $0$ if there is no edge between these nodes. The network having binary values for edges is called an \textit{unweighted} networks  If the matrix is symmetric, namely $A_{ij} = A_{ji},  \forall i,j \in V$, the matrix represents an \textit{undirected} network in which edges do not have directionality at all. If the matrix is asymmetric, on the other hand, it's representing a \textit{directed} network, where an element $A_{ij}$ implies an edge originating from the node $i$ and pointing to the node $j$. If edges of the network are weighted, meaning that edges can have real values, then the network is called \textit{weighted}. 
	
	Related with weighted networks is \textit{multigraph} in which any pair of nodes is allowed to have multiple edges connecting them. One should be careful here with the representation with an adjacency matrix for weighted network and multigraph, as the interpretation of the value of $A_{ij}$ depends on the context in which one treats the graph for the problem. For example, $A_{ij} = 2$ may describe an edge between $i$ and $j$ weighted by $2$, or it may represent \textit{two} edges running between $i$ and $j$. If diagonal elements of a matrix $A$ are non-zero, the network has \textit{self-loops} which indicates that there are edges originating from and pointing to the exact same node. In many studies such self-loops are simply ignored for the sake of simplicity.
	
	The last, but not the least kind of network is \textit{bipartite network}, in which there are two groups of nodes and edges exist \textit{only} between nodes of different groups. An example of bipartite network is a network of cooperate board membership where there are groups of companies and board members and edges connect companies and board members. Often times, a bipartite network is converted into a network in which there is only one kind of nodes, for instance only board members in the case of cooperate board membership network, by an operation called one-mode projection. In the one-mode projected network of board members, an edge between persons $i$ and $j$ now represents the frequency that they sit on the same board for companies.
	
	As such, there are many kinds of networks, each describing networks differently. In our study, for the sake of simplicity and comparability, we focus on using a \textit{simple graph}, in which edges have no directionality and weight and there is no self-loops. A set of procedures for converting non-simple graphs into simple graphs is defined as follows:
	
	\begin{enumerate}
		\item if a network is directed, then discard the directionality.
		\item If a network is weighted, then convert any non-zero weight into $1$ and $0$ otherwise.
		\item If a network is multigraph, namely having multiple edges between a pair of nodes, then merge them into one edge.
		\item if a network contains self-loops, then discard them.
	\end{enumerate}

	\subsection*{Structural features of a network}
	There are many ways to characterize a network in a quantitative manner that use the network's structural features \cite{Newman:NetworksIntro,NetworkCharacterizationSurvey}. Some network features, however, are implicitly correlated with the size of networks, which itself is a very strong feature: the number of nodes in web graphs is usually a magnitude of $10^6$ or more whereas ecological food webs contain usually less than 100 nodes. Examples of such features include, for instance, distance-related features such as the average path length and network diameter that are believed to grow approximately in $O(\log n)$, where $n$ is the number of nodes in a network. Another structural feature correlated with the size of the network is the modularity of network \cite{Modularity1}. Simply put, modularity quantifies the degree to which how much an observed network can be partitioned into modules within which edges are densely present but between which edges are sparsely present. The derivation of modularity is essentially based on comparing the original network with a random network having the same degree distribution which is called the configuration model and the value of modularity is practically the degree of difference between the original graph and its randomized counterpart. If the size of a network is large, the random network becomes so different that the modularity value becomes large as well \cite{ResolutionLimit, ModularityLimit}.
	
	The feature set we use in this study is scale-invariant, meaning that the size of network does not affect the value of a feature. This set of features allows us to compare networks without the notion of network size. In the following sections we describe important structural features that are relevant in this study.	
	 
\begin{table}[h!]
  \begin{center}
    \begin{tabular}{ p{3cm} | p{5cm} } 
      Name of the feature & Explanation  \\ 
      \hline
      Clustering coefficient &  The probability that a connected triplet ($k=3$ subgraph) is a triangle \\  
      Degree assortativity &  Correlation between a pair of connected nodes' degree. \\  
      Network motifs & The normalized z score of a subgraph's frequency compared to that of an ensemble of random networks having the exact same degree distribution.
    \end{tabular}
    \caption{Scale-independent network features.}
    \label{tab:feature}
  \end{center}
\end{table}

	 \medskip \paragraph*{Degree.}
	The \textit{degree} of a node in a network is the number of edges attached to the node. For a node $i$ in an unweighted network, the degree $k_i$ of the node can be written mathematically as:
	
	\begin{equation}
 	 k_i = \sum_{j = 1}^n A_{ij}.
	\end{equation}

\medskip \paragraph*{Clustering coefficient.}
	Clustering coefficient, which describes a degree of \textit{transitivity} in a network, is one of the most widely used metrics in network analysis especially in the context of social network. Transitivity in the context of networks means that if nodes $a$ and $b$ are connected as well as nodes $b$ and $c$, then nodes $a$ and $c$ are connected. Mathematically, the definition of clustering coefficient of a network is given by the following equation:
	\begin{equation}
	C = \frac{\text{number of closed paths of length two}}{\text{number of paths of length two}}.
	\end{equation}

\medskip \paragraph*{Degree assortativity.}
	\textit{Assortativity} in a network indicates a tendency for nodes to be connected to other nodes with similar node attribute. In a social network, for instance, node attributes which could be the basis for assortativity include language, age, income, alma mater, etc. On the other hand, \textit{disassortativity} exhibits nodes' tendency to be connected to other nodes having different node attribute in the network. An example network of disassortatvity would be a social network of heterosexual-relationship among people. 
	\textit{Degree assortativity} is a form of assortativity in which nodes with similar degree value tend to be connected together. Therefore in a network exhibiting high degree assortativity, there is a core of highly connected nodes with the high degree and periphery of nodes sparsely connected to other low-degree nodes.
	
	Here, we first define the covariance of $x_i$ and $x_j$ for the vertices at the ends of all edges, which leads us to a general \textit{assortativity coefficient}:
	
	\begin{equation}
		\text{cov}(x_i, x_j) = \frac{1}{2m}\sum_{ij}(A_{ij} - \frac{k_i k_j}{2m})x_i x_j,				
	\end{equation}
where $x_i$,$x_j$ are nodes $i$ and $j$'s attributes. This covariance will be positive if both $x_i$ and $x_j$ have, in average, similar values, and will be negative if both $x_i$ and $x_j$ tend to change in a different direction. We normalized cov($x_i$,$x_j$) by another quantity which represents a perfect assortativity, so that it takes a value $r \in [-1 ,1]$. The perfect matching happens if $x_i = x_j$ for all edges, and cov($x_i$,$x_i$) becomes:
	\begin{equation}
	\begin{split}
	 \text{cov}(x_i, x_i) &=  \frac{1}{2m}\sum_{ij}(A_{ij} x_i^2 - \frac{k_i k_j}{2m} x_i x_j) \\
	 		             &=  \frac{1}{2m}\sum_{ij}(k_i \delta_{ij} - \frac{k_i k_j}{2m}) x_i x_j,
	\end{split}
	\end{equation}
	where $\delta_{ij}$ is Kronecker delta.
Thus, the normalized covariance becomes as follows:
	\begin{equation}
	\begin{split}
	 r = \frac{ \text{cov}(x_i, x_j)}{ \text{cov}(x_i, x_i)} = 
	 \cfrac{\sum_{ij}(A_{ij} - \cfrac{k_i k_j}{2m})x_i x_j}{\sum_{ij}(k_i \delta_{ij} - \cfrac{k_i k_j}{2m}) x_i x_j}.
	\end{split}
	\end{equation}

Degree assortativity coefficient can easily be obtained by substituting $x_i$ and $x_j$ with degrees of respective vertices, thus:
	\begin{equation}
	 r =  \cfrac{\sum_{ij}(A_{ij} - \cfrac{k_i k_j}{2m})k_i k_j}{\sum_{ij}(k_i \delta_{ij} - \cfrac{k_i k_j}{2m}) k_i k_j}.
	\end{equation}

\medskip \paragraph*{Network motifs.} 
The idea of \textit{network motifs} was first introduced by Milo \textit{et al.} \cite{Milo_motif}. A network motif is a sub-graph of a network that appears more statistically significant than in random networks having the same degree distribution, namely the configuration model. There are a number of studies using network motifs of directed networks, where edges have directions, especially in biological sciences \cite{Alon2007, MotifsInBrain, NetworkMotifsEcoli}. Biologists have been particularly interested in motifs of networks, such as gene regulatory networks, since they are thought to correspond to the functional building blocks in biological systems and may help scientists understand the underlying principles of biological complex systems.

\begin{figure}[t!]
	\begin{center}
		\vspace{0.5cm}
		\includegraphics[clip,width=7cm,height = 2cm]{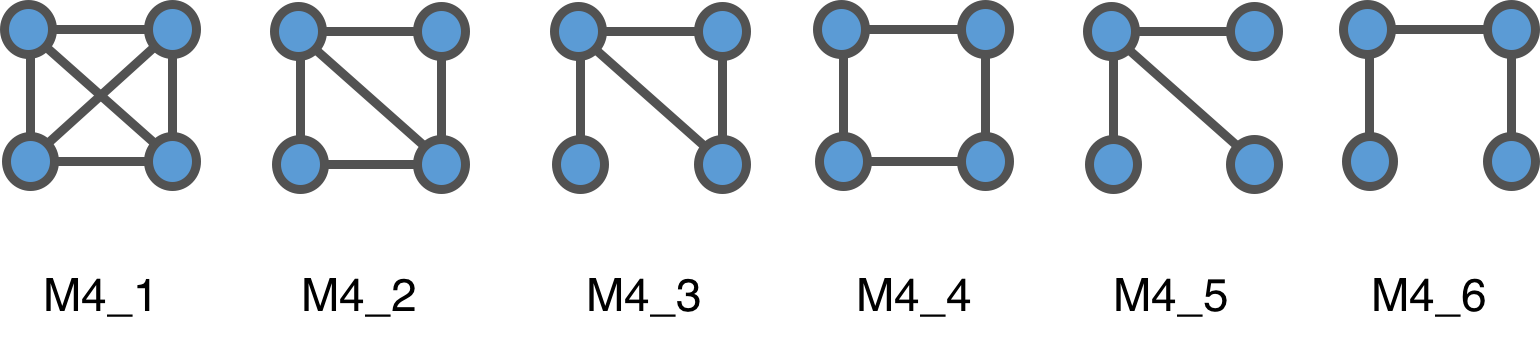}
		\vspace{0.5cm}
		\caption{Undirected, non-isomorphic subgraphs (network motifs) of $k=4$ nodes.}
		\label{motifs}
	\end{center}
\end{figure}

As opposed to the number of studies using directed network motifs, in this study we only use network motifs of undirected networks, that produce fewer variations in motif kinds, but can be applied to any network regardless of edge directionality which is crucial for our study. Fig.~\ref{motifs} shows the complete list of $k=4$ undirected connected subgraphs used in this study.

In order to quantify network motifs, we first count the occurrence of each sub-graph in the original network, then repeat the same process on the configuration models. After counting occurrences of sub-graphs in both original and multiple random networks, we proceed to calculate z-score for each sub-graph $i$ as follows:
	\begin{equation}
	Z_i = \cfrac{N_i^{\text{original}} - \langle N_i^{\text{random}} \rangle }{\sigma_i^{\text{random}}},
	\end{equation}

where $N_i^{\text{original}}$ is the number of occurrence of a sub-graph $i$ in the original network and $ \langle N_i^{\text{random}} \rangle$ and $\sigma_i^{\text{random}}$ are the average and the standard deviation of the number of occurrence of a sub-graph $i$ in an ensemble of random networks. It is usually convenient to normalize this z-score as some networks exhibits very large values due to the size of the networks. Such normalized z-score is called \textit{significance profile} and is defined as follows:
	\begin{equation}
	SP_i = \cfrac{Z_i}{\sum_j Z_j^2}.
	\end{equation}
	
In this study, we treat the significance profile for each subgraph as a feature and in total we have six features as network motifs.
\newline

\section*{Network data sets}
	Our real-world network data sets are drawn from the Colorado Index of Complex Networks (ICON) \url{icon.colorado.edu}, which is a large index of real-world network data sets available online. ICON is an index not than a repository, and hence we used ICON to locate the original network data files associated with a large number of studies from a diverse set of scientific domains. 
	
	Since the data format of real world networks is not standardized, we proceeded to convert all the data into a single format called \textit{Graph Modeling Language} (GML) \cite{GML}. The format allows us to flexibly specify arbitrary node and edge attributes. We, however, do not use any node and edge attribute, including edge weight, as well as edge directionality at all since not all networks have such properties and we wanted to analyze a diverse set of networks. Thus, we treat all networks used in this study as a simple graph which is defined in the previous section. We have used only a fraction of all networks available on ICON due to the time constraint. 
	
	We have also added some synthesized network data which are generated from four specific models briefly explained in the following:
	 
	\begin{enumerate}
		\item Erd\H{o}s-R\'enyi random network model (ER Network) \cite{ER_Network}, where given $n$ the number of edges and $p$ the probability that a pair of nodes gets connected, for each pair of nodes in the network, one connects the nodes according to $p$. This model yields the average path length of $O(\log n)$ and low clustering coefficient.
		
		\item Watts-Strogatz model (Small World) \cite{watts1998cds} which produces a network having the high clustering coefficient and low average path length of $O(\log n)$. The model starts with a grid network and then rewires some edges according to some probability $p$. The rewiring makes the network's path length smaller while keeping the high clustering due to the gird structure.
		
		\item Barab\'asi-Albert model (Scale Free) \cite{Barabasi99emergenceScaling} with which one grows a network over the course of time. Newly added nodes have a fixed number of edges attached to them, and these edges connect to the existing nodes according to the probability $p$ that is proportional to the degree of an existing node. Therefore nodes having many connections will be more likely to receive more edges attached to them. Although this model was originally invented by Price in a paper in 1965 \cite{deSollaPrice1965}, we call the model as BA model since it's more widely known as its name.
		
		\item The Forest Fire network model (Forest Fire Network) \cite{ForestFire} which is a network generative model with the following procedures: (i) a newly added node $u$ attaches to (cites) some existing nodes, called \textit{ambassadors}, chosen uniformly at random; (ii) for each newly cited node $v$, its incoming and out-going neighbors are also cited by $u$, the new node; (iii) the same procedure is done recursively for all of the newly cited nodes.
	\end{enumerate}

\begin{figure*}
\centering 
\subfloat[The categorical ratio of network domains]{%
  \includegraphics[width=0.85\columnwidth]{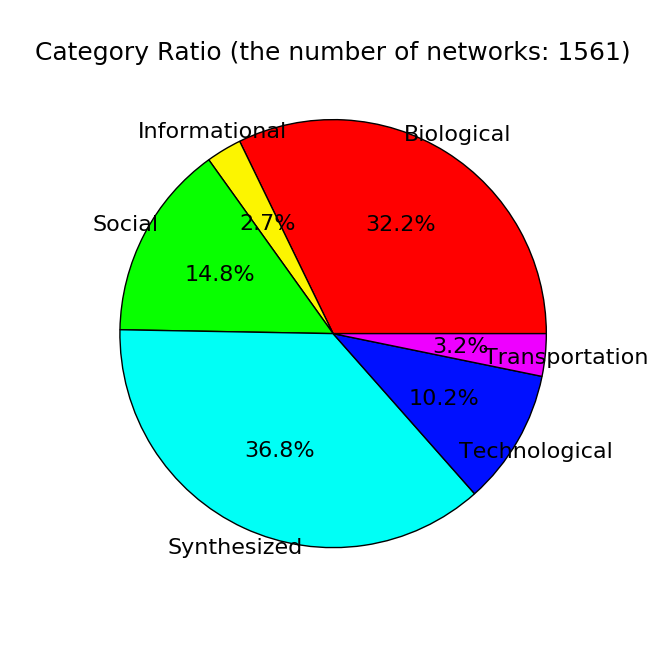}%
  \label{domain_ratio}%
}\qquad
\subfloat[Count distribution of network sub-domains.]{%
  \includegraphics[width=1.10\columnwidth]{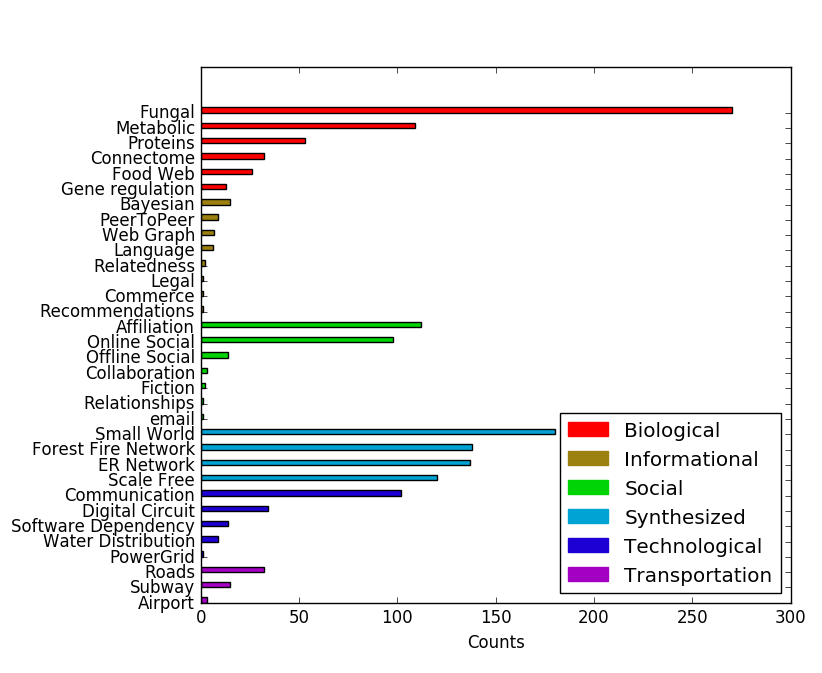}%
  \label{sub_dist}%
}

\caption{ Sub-domains of the same network domain are grouped together having the same color in the figure. Color code from top: Biological, Informational, Social, Synthesized, Technological and Transportation.} \label{category_dist}.
\end{figure*}

We show all the details of network sub-domains for our study in the Table \ref{tab:subdomain} (located at the end of this paper). The distributions of network domains and sub-domains, as shown in Figs.~\ref{domain_ratio} and \ref{sub_dist}, are very skewed since instances of some network categories are hard to obtain due to their inherent difficulty of collecting data or legal concerns, or hard to analyze due to their network size and this leads us to explore several sampling methods, which are explained in the following section. 
	
After converting into GML format, we calculate a set of features explained in the previous section for each network. We have extensively used a Python library \textit{igraph} \cite{igraph} for extracting features including clustering coefficient and degree assortativity and other miscellaneous operations on network data. For calculating network motifs, we used a parallel motif computing algorithm for undirected motifs developed by Ahmed \textit{et al.} \cite{ahmed2015icdm}. A number of computations involved in this study are parallelized by using a command-line tool \textit{GNU Parallel} \cite{GNUParallel}.

\section*{Methodological issues}
After having converted networks into vectors in the feature space, there are a number of possible ways to analyze the distribution of points and labels and possibly learn the concept that governs such distribution in the feature space. In this study, we use random forest classifier along with the confusion matrix as a way to learn the underlying concept that differentiates different classes of networks. As we have seen in the previous section, the distribution of class labels is obviously skewed which leads us to use several sampling methods that are supposed to alleviate the problem. Here, we explain such methodologies in detail.

	\subsection*{Managing class imbalance}
Most of the machine learning algorithms perform well on evenly populated instances of multiple classes. However, once this class balance no longer persists, the algorithms perform poorly on minority classes. This problem, called \textit{class imbalance}, causes any machine learning algorithm that is naive to the data set to focus exclusively on the majority class, ignoring any minority classes. One of the most widely used approaches for mitigating the problem is sampling the data set of interest so that the distribution of classes becomes balanced. Although there are many proposed sampling strategies as of now \cite{SurveySampling}, we primarily use three sampling strategies in this study: random over-sampling, random under-sampling and SMOTE \cite{SMOTE}.

 Here we establish the mathematical notations used in explaining sampling methods. Let $S$ be a set of pairs of $\vec{x_i}$ and $y_i$, namely $S =\{(\vec{x_i},y_i)\}$, for $i = 1,...,n$ where $n$ is the number of data, $\vec{x_i} \in X$ is an instance of networks in the $N$-dimensional feature space and $y_i \in Y = \{1,...,C\}$ is a class label associated with the instance $\vec{x_i}$.

		\medskip \paragraph*{Random over-sampling.}
		This method is one of the simplest strategies for mitigating the class imbalance problem. It over-samples any minority classes to an extent that the number of instances in each class becomes even. Here we explain this sampling method in a mathematical sense. Let $S_{\rm{maj}} \subset S$ be the majority class, meaning a class having the largest number of instances, $S_{\rm{min}}^{j} \subset S$ the $j$th minority class where $j = 1,...,C-1$ and $P$ a set $\{S_{\rm{maj}}, S_{\rm{min}}^{1},...,S_{\rm{min}}^{C-1} \}$ such that
	
	\begin{equation}
	S_{\rm{maj}} \cup S_{\rm{min}}^{1} \cup S_{\rm{min}}^{2},..., \cup\, S_{\rm{min}}^{C-1} = S,
	\end{equation}
	and
	\begin{equation}
	\forall S_1,S_2 \in P \land S_1 \neq S_2 \Rightarrow S_1 \cap S_2 = \emptyset,
	\end{equation}

where $\emptyset$ means an empty set. Let $E_{\rm{min}}^j$ be a set of points that are sampled at uniformly random from a set $S_{\rm{min}}^{j}$ that satisfies the following equality:

	\begin{equation}
	|S_{\rm{min}}| + | E_{\rm{min}}^j | = |S_{\rm{maj}}|.
	\end{equation}
	
 We then append the set $E_{\rm{min}}^j$ to the corresponding set of the minority class $S_{\rm{min}}^{j}$, namely $S_{\rm{min}}^{j} := S_{\rm{min}}^{j} + E_{\rm{min}}^j$. The drawback of random over-sampling is a potential poor generalization due to the overfit of a classifier to the over-sampled instances.
	
	\medskip \paragraph*{Random Under-sampling.}
	On the other hand, this method under-samples the majority class, essentially throwing out some data in order to make the ``cloud'' of data points sparser. This``throwing out instances'' implies an obvious drawback of this sampling method: It discards potentially important instances that compose the backbone of majority classes, implying the true shape of majority class is no longer retained. Here, again, we explain this method using mathematics. Let $S_{\rm{min}} \subset S$ be the minority class, meaning a class having the least number of instances, $S_{\rm{maj}}^{j}$ the $j$th majority class where $j = 1,...,C-1$ and $P$ a set $\{S_{\rm{min}}, S_{\rm{maj}}^{1},...,S_{\rm{maj}}^{C-1} \}$ such that

	\begin{equation}
	S_{\rm{min}} \cup S_{\rm{maj}}^{1} \cup S_{\rm{maj}}^{2},..., \cup\, S_{\rm{maj}}^{C-1} = S
	\end{equation}
	and
	\begin{equation}
	\forall S_1,S_2 \in P \land S_1 \neq S_2 \Rightarrow S_1 \cap S_2 = \emptyset.
	\end{equation}
	
Let $E_{\rm{maj}}^j$ be a set of points that are sampled at uniformly random from a set $S_{\rm{maj}}^{j}$ that satisfies the following equality:

	\begin{equation}
	|S_{\rm{min}}| = | S_{\rm{maj}}^j -  E_{\rm{maj}}^j |.
	\end{equation}
Then, we subtract the set $S_{\rm{maj}}^j$ by $E_{\rm{maj}}^j$, thus $S_{\rm{maj}}^j := S_{\rm{maj}}^j -  E_{\rm{maj}}^j$.
		
	\bigskip \paragraph*{SMOTE.}
		\textit{Synthetic Minority Over-sampling Technique}, widely known as SMOTE \cite{SMOTE}, is an alternative sampling method that synthesizes data points in the training set for a classifier. Mathematically speaking, SMOTE is quite similar to random over-sampling, except for how it generates the set $E_{\rm{min}}^j$ for the minority class $S_{\rm{min}}^j$. For each set of the minority class $j$, namely $S_{\rm{min}}^j$, consider $K$ nearest neighbors of a point $\vec{x_i} \in S_{\rm{min}}^j$ in the $N$-dimensional feature space. To generate a new point, first pick up one of the $K$ nearest neighbors, say $\vec{x_{n}}$, then multiply the corresponding feature vector difference with a weight  $\delta$  chosen from an interval $ [0,1]$ at uniformly random and add this vector to $\vec{x_i}$. Therefore, we have a newly synthesized point defined as follows:
		
		\begin{equation}
		\vec{x_{\rm{new}}} = \vec{x_{i}} + \delta*(\vec{x_{n}} - \vec{x_{i}}).
		\end{equation}

\begin{figure}[t!]
		\begin{center}
		\vspace{0.5cm}
		\includegraphics[clip,width=7.5cm,height = 2cm]{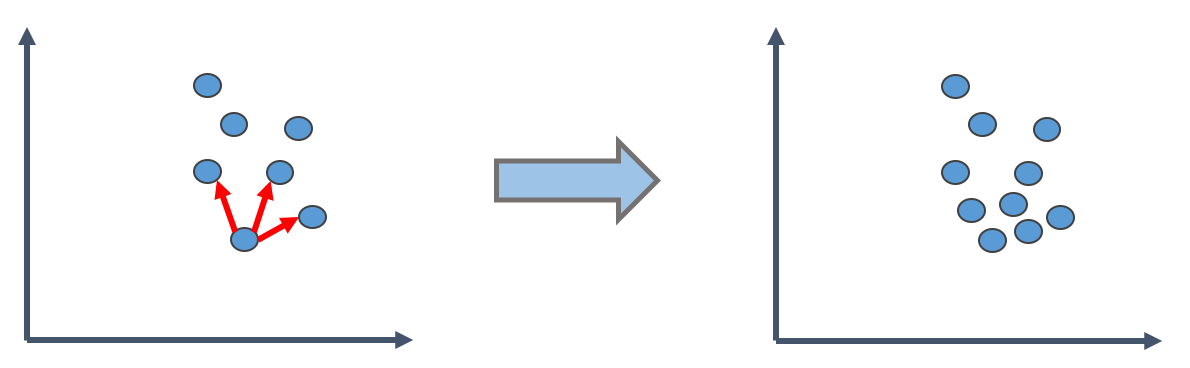}
		\vspace{0.5cm}
		\caption{Synthesizing phase of SMOTE. Here the number of nearest neighbors $k$ is 3.}
		\label{smote}
		\end{center}
		\end{figure}

And one repeats this procedure for other neighbors of the point $\vec{x_i}$ and construct a set $E_{\rm{min}}^j$. See the figure \ref{smote} for visualization of the synthesizing process.	
		
The core concept of SMOTE is filling out the cloud of minority class instances by interpolating existing data points so that it closely resembles a \textit{convex set}. This idea, making a convex set of minority instances by interpolation, assumes that the shape of a manifold of the underling data distribution itself is convex. In a high dimensional space, it is often the case that the distribution of data forms a quite intricate non-convex manifold on which the data we observe is generated. Therefore one must be careful when using a sampling method like SMOTE that the resulting set of points by interpolation of data points of such complex manifold is likely to be a convex set, that may radically be different from the underlying concept that a classifier tries to learn.

	\subsection*{Learning what distinguishes domains}
	In order to find similarities and differences of data of different classes, one needs to develop a notion of similarity between the classes. In this study we derive such notion of similarity from confusion matrices that are produced by random forest classifiers. Following sub-sections describe the details of decision tree which is an essential component of random forest, random forest classifier and confusion matrix.
	
	\bigskip \paragraph*{Decision trees.}
		Decision tree is a model which describes the relationship between input variables and output class by recursively asking a question on a single input variable and splitting the data set into two based on the answer to the question until a data set has enough homogeneity of a class in it. In a high dimensional space, such spitting the data set corresponds to hyperplane in the space. The algorithm for learning decision tree splits the data set based on a criterion of values of an input variable such that the resulting data sets become less heterogeneous or less impure in terms of class labels. One of the widely used such criteria and the one we use in this study is \textit{Gini impurity}. The definition of Gini impurity for a data set with $J$ classes is the following:
	\begin{align}
	I_G(f) & = \sum_{i=1}^J f_i(1-f_i) = \sum_{i=1}^J (f_i-f_i^2) \nonumber \\
	  & = \sum_{i=1}^J f_i - \sum_{i=1}^J f_i^2 = 1- \sum_{i=1}^J f_i^2,
	\end{align}
where $f_i$ is a probability that an item that belongs to class $i$ is chosen in the data set. Gini impurity becomes $0$ if all items in a set belong to the same class, meaning the set is ``pure'' and takes a value greater than $0$ if the set contains items of multiple classes. Each splitting essentially seeks the best possible value of an input variable such that the decrease of Gini impurity is the largest when the data set is split at the value (or hyperplane defined with it). Splitting continues until no further improvement can be made and the terminal of a tree are called leaves of the tree, each corresponding to one of the class labels in the data set.

	\bigskip \paragraph*{Random forests.}
The popular random forest classifier is a type of \textit{ensemble learning method} that combines a number of so called```sweak'' classifiers together \cite{RandomForest}. When it's given a data set to predict after training weak classifiers, it outputs the majority of all outputs from the weak classifiers and this aggregation of weak classifiers prevents random forest from overfitting to the training data. In random forest, such weak classifiers are a decision tree which is explained in the previous section.

The learning phase of random forest classifier first involves random sub-sampling of the original data with replacement for $B$ times, each time the sub-sampled data set is fed to a decision tree. For each decision tree a set of randomly sampled input variables (features) is used for splitting and this random selection of features is called random subspace method or feature bagging. This prevents the classifier to focus too much on feature sets that are highly predictive in the training set. 

One of the advantageous byproducts of random forest is that one can rank input variables or features based on the importance in the classification. Each time a split is made on a node in a decision tree, the decease of Gini impurity can be attributed to selecting a feature as a splitting point.  Calculating the average decrease of Gini impurity for selecting a feature over all decision trees in random forest gives us the importance of the feature that is very consistent with the result of the original method for calculating variable importance \cite{RandomForest,RandomForestOnline}. This ranking of feature importance is the crucial part of the analysis that we describe in the next section.

	\bigskip \paragraph*{Confusion matrices and Domain similarity.}
	A confusion matrix depicts when and how frequently a classifier makes mistakes. The row labels of the matrix usually correspond to  true labels and column labels correspond to predicted labels. An element $c_{ij}$ in a confusion matrix represents the number of occurrences that a classifier predicted an instance of class $i$ as class $j$. So it is easy to notice that diagonal elements of a confusion matrix, namely $c_{ii}$ for $i = 1,...,n$ represents the correct predictions of a classifier. What we are interested in, however, lies in off-diagonal elements of a confusion matrix. The information that a classifier gets confused with classes $i$ and $j$ implies the similarity between classes: if the points of two different classes in a high dimensional feature space are often misclassified as each other, the points of the two classes are in fact close to each other in the feature space, implying that these classes are so similar that a classifier cannot distinguish one from the other. The information in a confusion matrix, when and how frequently a classifier makes mistakes, is essential in order for us to answer one of the questions we have asked:
	\begin{center}
	\textit{Are there any sets of network categories that are inherently indistinguishable from each other based solely on network structure?}
	\end{center}

It is tempting to use a confusion matrix as a similarity matrix owing to the fact that off-diagonal elements imply the similarities among class labels. This, using a confusion matrix as a similarity matrix, involves following issues however:
\begin{enumerate}
	\item In a confusion matrix, classes with abundant data tend to have large counts for elements in the matrix due to the abundance of test data whereas classes with fewer data have fewer counts in the matrix.
	\item Usually the confusion matrix is not symmetrical, but a number of similarity-related methods assume an input matrix has the symmetry.
\end{enumerate}

Therefore we proceed on the following operations in order to derive a similarity matrix based on a confusion matrix:
\begin{enumerate}
	\item Normalize each row $i$ of the confusion matrix so that $\sum_{j=1}^J c_{ij} = 1$
	\item Symmetrize the resulting matrix from operation $1$ by setting each pair of symmetric elements as: $c_{ij},c_{ji} = \max (c_{ij},c_{ji})$
\end{enumerate}

\section*{Experimental results}
In this section, we present experimental settings for analyses, each of which helps us to gain insights for answering the questions we proposed in the introduction, namely questions about the distinguishing structural features, existence of indistinguishable pairs of network categories and its implication. We present the results of such analyses in a sequential order along with those questions. 

\subsection*{Discriminative features}
The first question of focus is about the set of distinguishing features that make a category of networks ``stand out'' among others. In order to address this question, we look at the statistics of feature importance that are derived from a random forest classifier. The assumption here is that distinguishing features should be able to split the data in the feature space into sub-spaces such that separation of class labels is good, meaning that these sub-spaces should only contain nearly a single class label within themselves.

The degree of goodness of separation is expressed as the decrease of Gini impurity and if a feature is distinguishing one category of networks from the others, then selecting the feature in a decision tree should gain a large decrease of Gini impurity in splitting the data and the feature should be highly ranked in the feature importance ranking. Therefore, if we observe a feature that ranks as the top in the ranking many times for binary classifications in which positive label corresponds to the class of interest and the negative label corresponds to everything else grouped together, we could assert that the feature is distinguishing for a specific kind of networks from the rest. 
 
\medskip \paragraph*{Individual feature approach.} First, since not all of the sub-domains have enough number of samples to support our analysis and due to the limitation of time and space, we select a set of six representative network sub-domains that have been investigated in a number of studies and attract a particular interest in each domain. These representative sub-domains are: protein interaction, ecological food web, metabolic, connectome (brain networks),  online social and communication (autonomous systems). 

Second, for each representative class we proceed to run binary classification 1000 times using random forest in which the sub-domain of interest corresponds to the positive and the other sub-domains grouped together correspond to the negative.  A set of features for this task includes: clustering coefficient, degree assortativity and $k = 4$ connected network motifs, thus the total of eight features. In each run we split the data set into training and test sets with the ratio of $7:3$ while preserving the ratio of class distribution. In each run the score of AUC (Area Under the ROC Curve) is calculated in order to see the performance of random forest classifier for the binary classifications. We also record for each run the ranking of feature importance in which features are sorted according to their Gini impurity decrease in the training phase of random forest classifier. We then average AUC scores over 1000 runs and aggregate all of the recorded rankings of feature importance.

Finally, we select the two most important features from the aggregated ranking, plot all of the data points in the two dimensional feature space in which $x$- and $y$-axes correspond to the most and second most important features, respectively. This visualization will help us understand to what degree and how a category of networks can be separated from the rest in the two dimensional feature space if there exists any pair of distinguishing features.

\begin{figure*}
\centering 

\subfloat[Protein interaction.  Average AUC score: 0.689]{%
  \includegraphics[width=0.625\columnwidth]{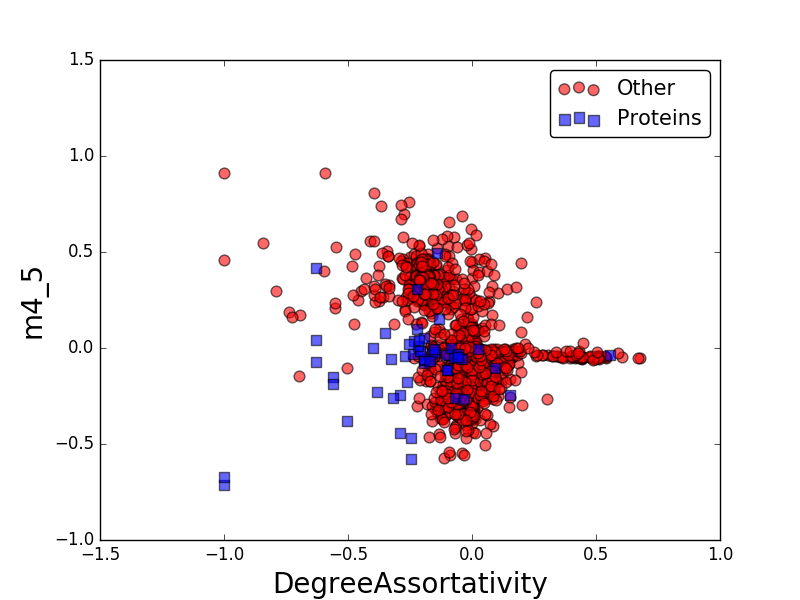}%
   \label{protein_2d}%
}\qquad
\subfloat[Ecological food web. Average AUC score: 0.804]{%
  \includegraphics[width=0.625\columnwidth]{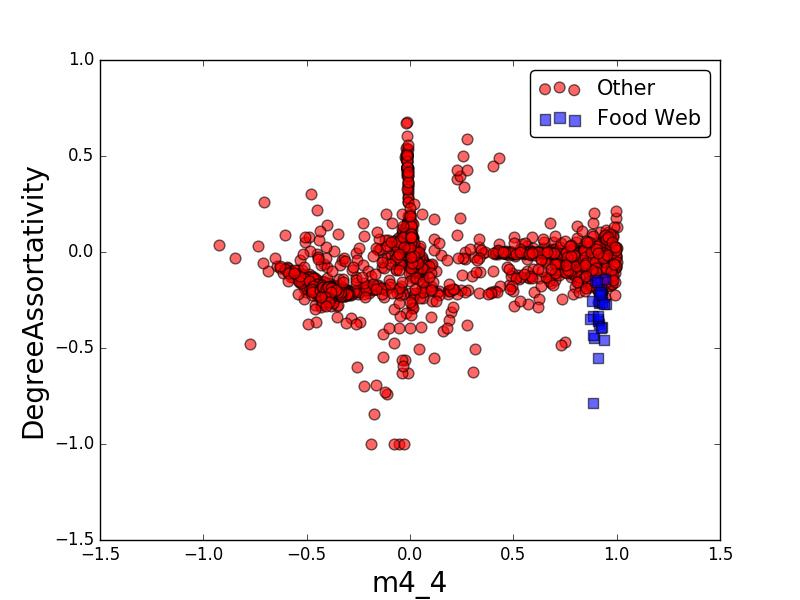}%
  \label{foodweb_2d}%
}\subfloat[Metabolic. Average AUC score: 0.94]{%
  \includegraphics[width=0.625\columnwidth]{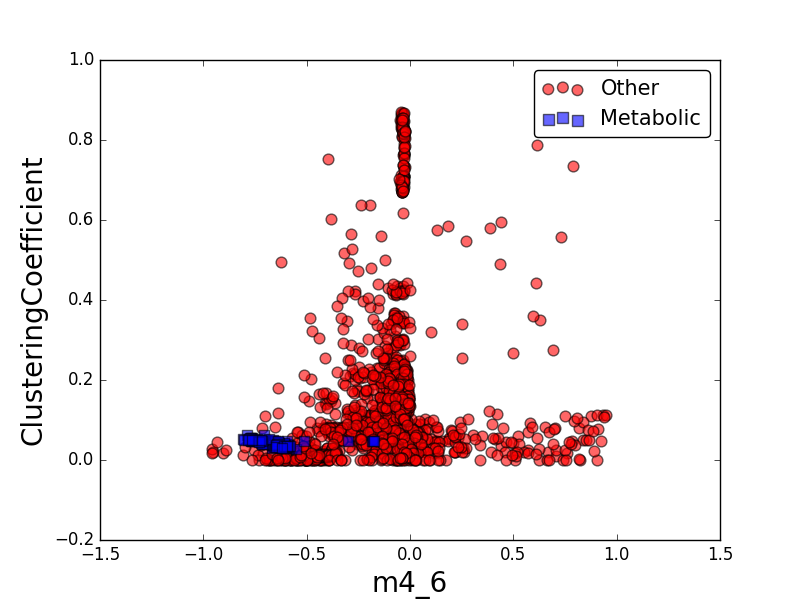}%
   \label{metabolic_2d}%
}\qquad

\medskip

\subfloat[Connectome. Average AUC score: 0.716]{%
  \includegraphics[width=0.625\columnwidth]{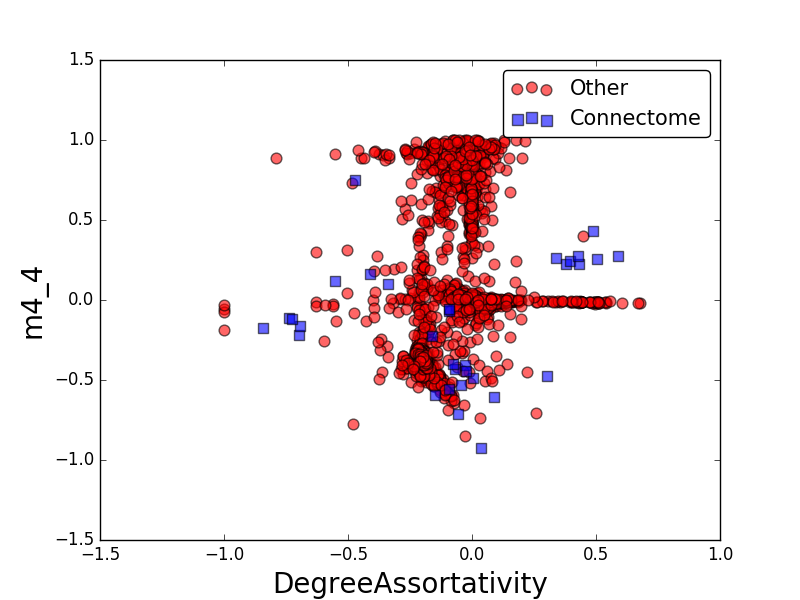}%
  \label{connectome_2d}%
}
\subfloat[Online social. Average AUC score: 0.94]{%
  \includegraphics[width=0.625\columnwidth]{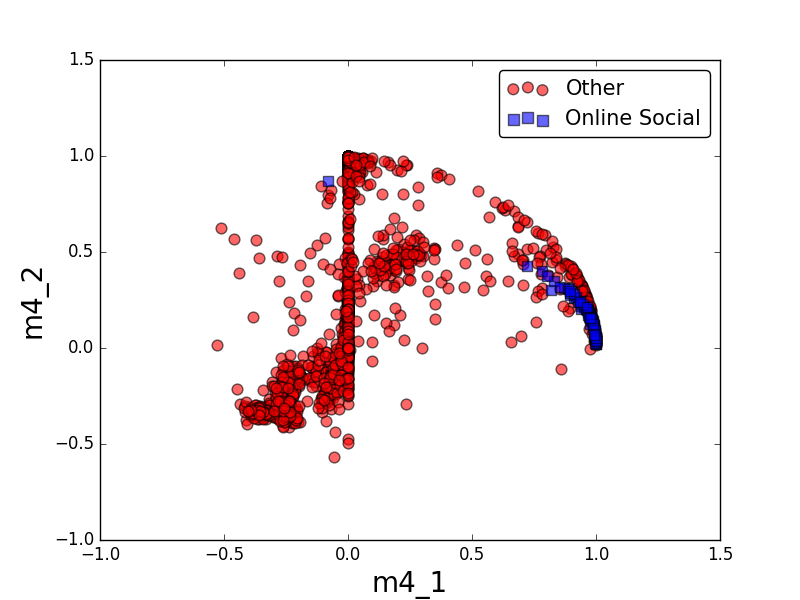}%
   \label{online_social_2d}%
}\qquad
\subfloat[Communication. Average AUC score: 0.992]{%
  \includegraphics[width=0.625\columnwidth]{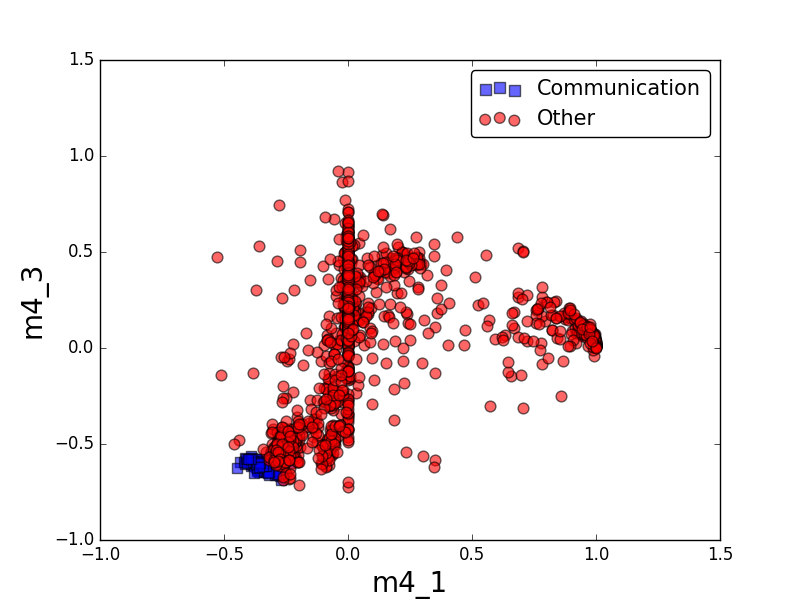}%
  \label{communication_2d}%
}

\caption{2D plots for all representative network sub-domains. The $x$-axis corresponds to the most importance feature and $y$-axis the second most important.} \label{2d_figures}.
\end{figure*}

\begin{figure*}
\centering 

\subfloat[Protein interaction. ]{%
  \includegraphics[width=0.625\columnwidth]{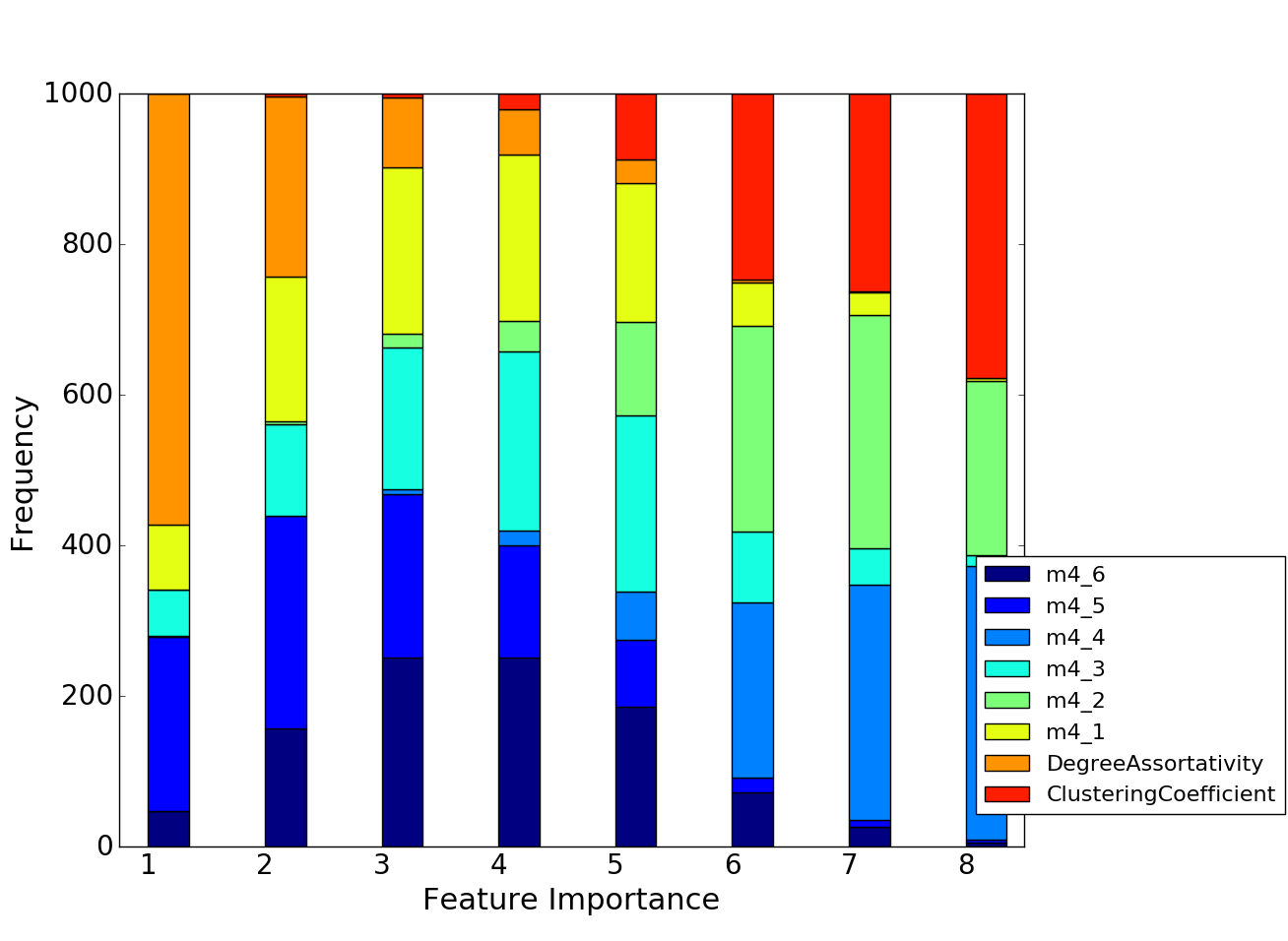}%
   \label{protein_feature}%
}\qquad
\subfloat[Ecological food web. ]{%
  \includegraphics[width=0.625\columnwidth]{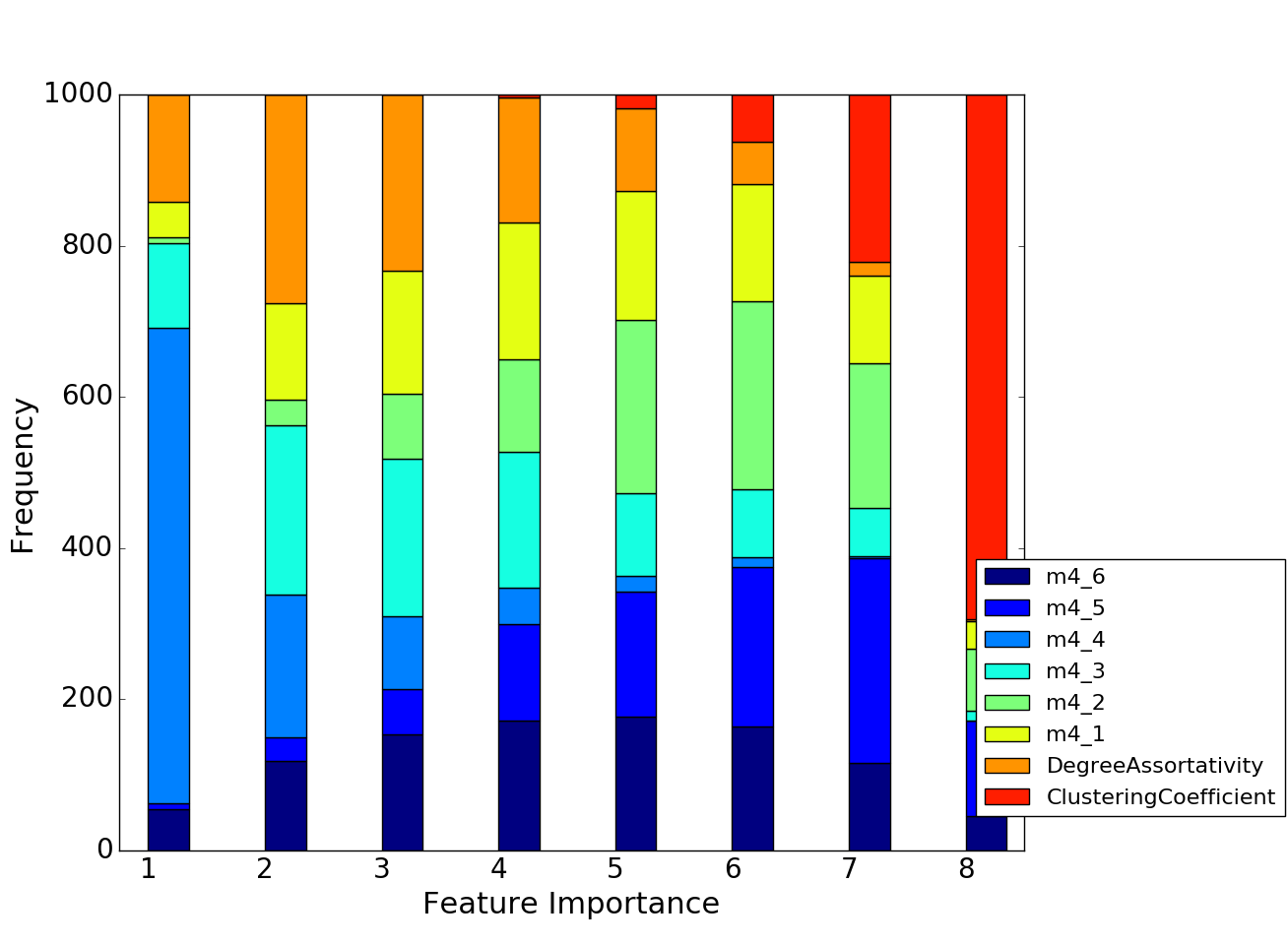}%
  \label{foodweb_feature}%
}\qquad
\subfloat[Metabolic.]{%
  \includegraphics[width=0.625\columnwidth]{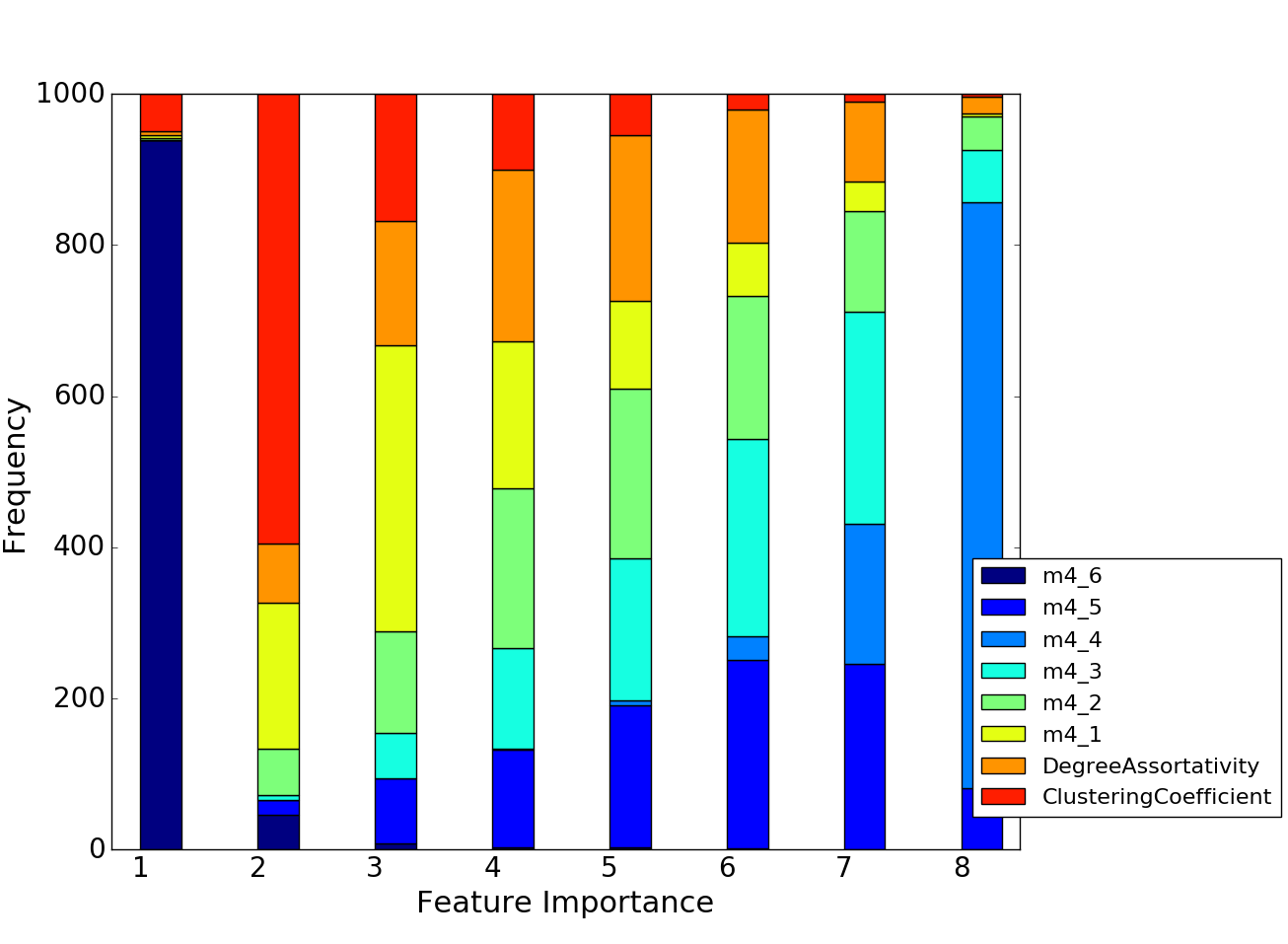}%
   \label{metabolic_feature}%
}

\medskip

\subfloat[Connectome.]{%
  \includegraphics[width=0.625\columnwidth]{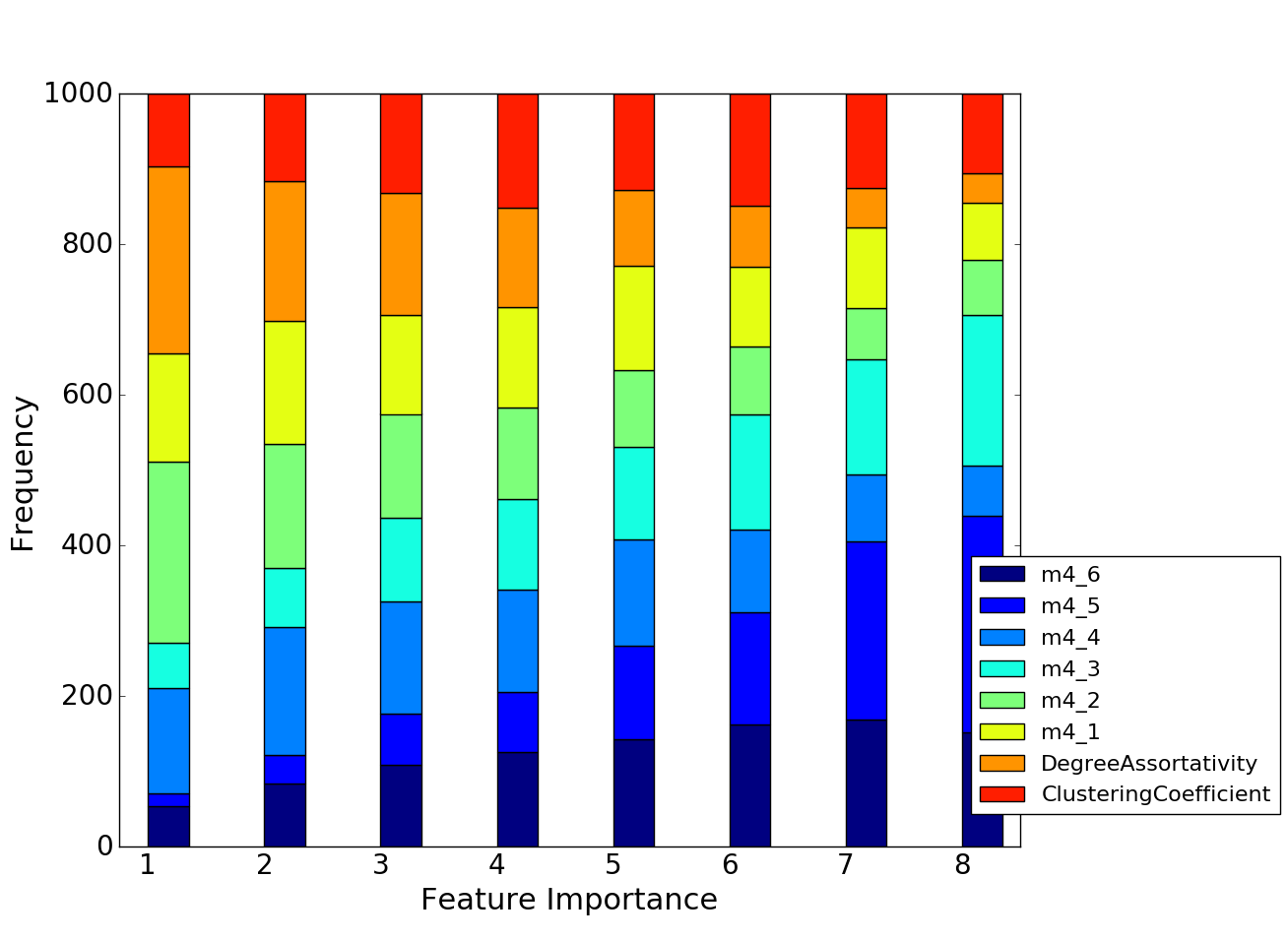}%
  \label{connectome_feature}%
}\qquad
\subfloat[Online social.]{%
  \includegraphics[width=0.625\columnwidth]{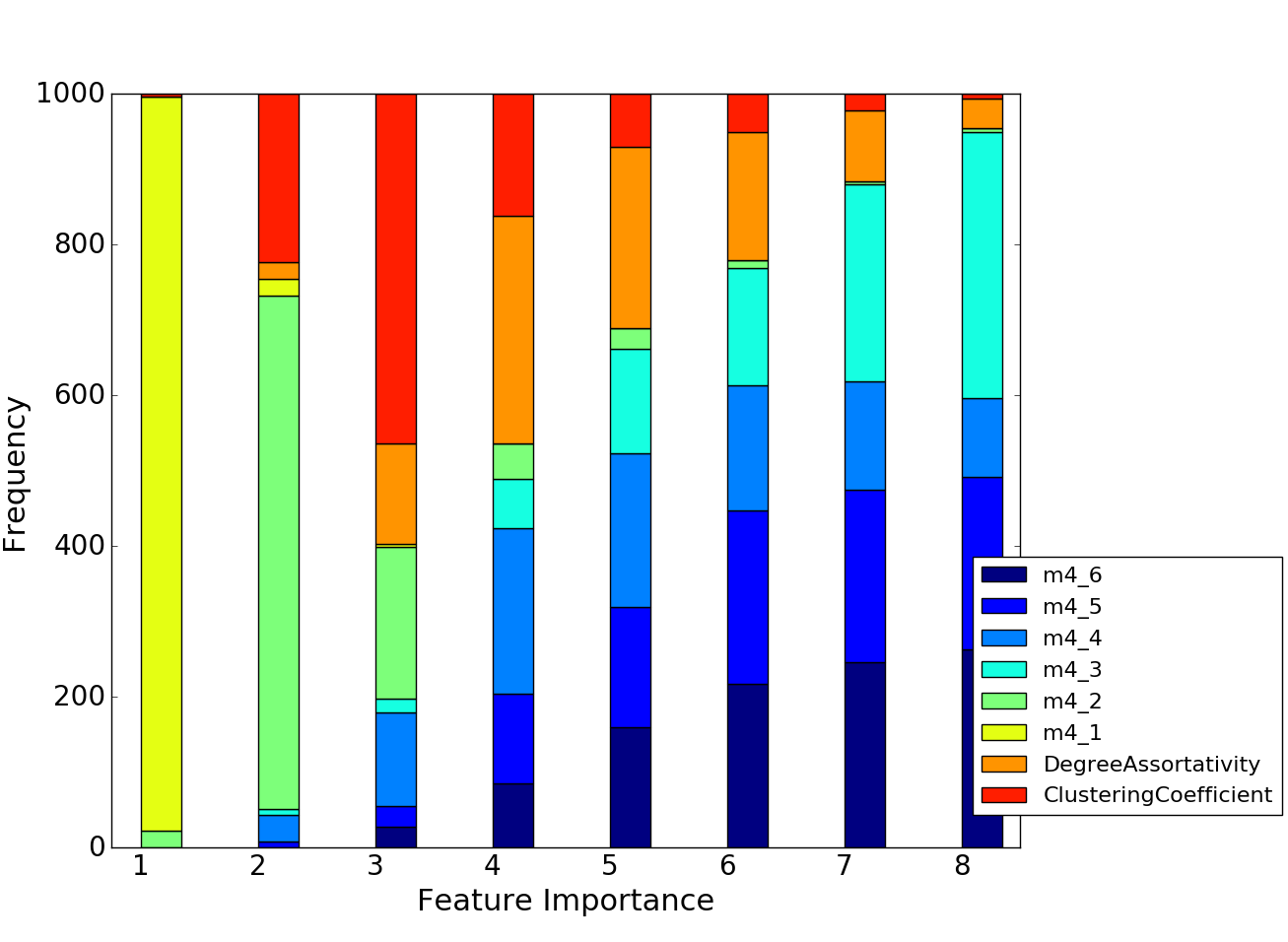}%
   \label{online_social_feature}%
}\qquad
\subfloat[Communication.]{%
  \includegraphics[width=0.625\columnwidth]{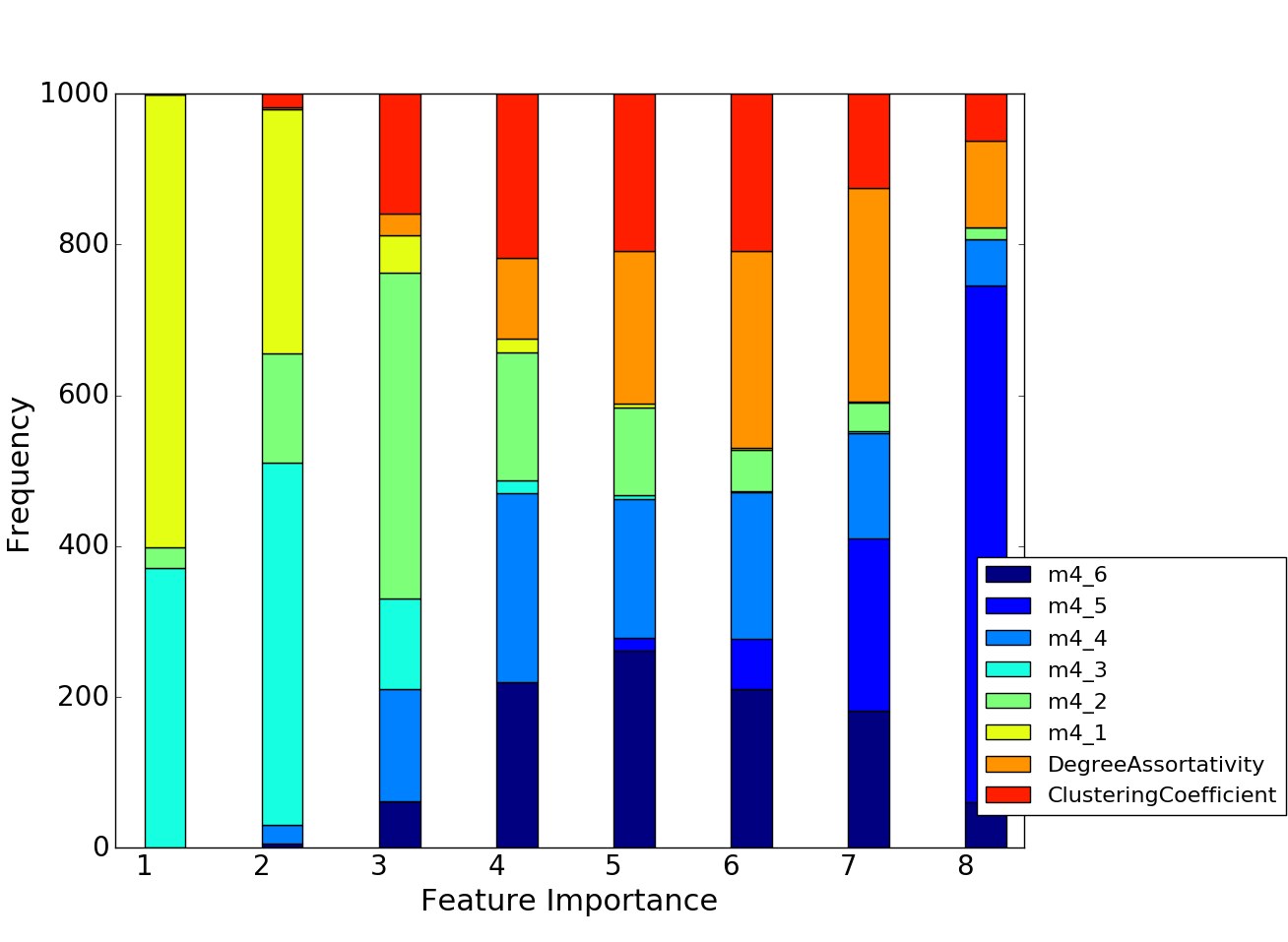}%
  \label{communication_feature}%
}

\caption{Aggregated rankings of feature importance. The height of a color bar indicates a frequency of the corresponding specific feature being at the rank. The importance decreases along the $x$-axis.} \label{feature_importance_figures}.
\end{figure*}

\medskip \paragraph*{Results on individual features.} 
Fig.~\ref{2d_figures} shows two dimensional plots for the representative network sub-domains with axes being the top two important features selected from the statistics of aggregated feature importance shown in Fig.~\ref{feature_importance_figures}. One important observation here is that the AUC score captures the separability and the degree of spread of a specific class label in the two dimensional feature space: the averaged AUC score for protein interaction networks shown in Fig.~\ref{protein_2d} is 0.693 and the data points are spread across the large region of the feature space; the 2D plot of communication networks as shown in Fig.~ \ref{communication_2d}, on the other hand, displays a cluster of points corresponding to the class and the averaged AUC score for binary classifications is 0.992 which is much higher than that of protein interaction networks. The degree of spread of data points implies the structural variability of networks within a single category. For example, protein interaction networks and connectome exhibit the high structural variability of networks in a space defined by the selected features, whereas other network categories such as metabolic networks, ecological food webs, etc. exhibit the low structural variability in the selected features. This suggests the possible need for adding a new set of features for protein interaction networks and connectome with which we could distinguish them from the other kinds.

Fig.~\ref{feature_importance_figures} shows the aggregated rankings of feature importance. In this figure, one can observe the general trend of an informative feature set and the strength of those features in the ranking. For example, the motif $m4\_6$, a connected path, of metabolic networks and the motif $m4\_1$, $k=4$ clique, of online social networks are the most important features and their strength is quite dominant:  $m4\_6$ of metabolic networks ranks as the first 939 times out of 1000 runs and $m4\_1$ of online social networks ranks as the first 974 times  out of 1000 runs, respectively. It is interesting to note that for online social networks clustering coefficient and degree assortativity, both long known for distinguishing features for social networks in general do not rank as the most or even the second most important features here.  On the other hand, the ranking of feature importance for connectome displays a lack of definitive features that can be seen in the almost equal frequencies of features appearing at any rank. This indicates that none of the features we have used cannot explain the network category succinctly: for online social networks we can describe the networks as having an extraordinarily number of $k=4$ cliques when compared to their randomized counterparts; we cannot, however, describe connectome  in such a way with the feature set.

These distinguishing features observed in the figures seem to have some implications for the process in a network. Ecological food-webs display the abundance of $m4\_4$ subgraphs, namely a square of four nodes, that are thought to be major constituent elements in the food-webs as it describes layers of the food chain \cite{Milo_motif,BiParallel}: animals in the same layer in the food chain do not often prey on each other, but prey on animals in a layer below and they are preyed on by animals in a layer above; usually this relationship is depicted in a directed network motif named as bi-parallel motif. Note that, however, this directed motif is converted into an undirected version of itself, namely $m4\_4$ in our study. Online social networks have an unusual number of $m4\_1$ subgraphs, namely $k = 4$ clique, and this indicates a strong local bonding mechanism in social networks: suppose a person $A$ has friends $B$ and $C$ that know each other (triangle of $A$,$B$ and $C$); if $B$ and $C$ have a mutual friend, called $D$, then it is likely in online social networks that the person $A$ is also a friend to $D$ and this forms a $k=4$ clique in the network.


On the other hand, communication networks namely the Internet at the level of autonomous systems exhibit the underrepresented number of  $m4\_1$ subgraphs. It was previously reported that the number of triangles in the communication networks is less than the expected value of the number of triangles for their randomized counterparts having the same degree distribution \cite{InternetClustering}. The $m4\_1$ subgraph, namely $k=4$ clique, contains four triangles within itself. Therefore it is not illogical to say that the underrepresentation of triangles approximately equals to the underrepresentation of $k=4$ clique, which leads us to claim that our finding coincides with the previous study. This underrepresentation of $k=4$ clique may imply the underlying growing or construction mechanism of communication systems: the whole system needs to be connected in order for the Internet to work, but does not need too many connections among autonomous systems as the cost for connections should be minimized. 

For some network sub-domains, however, implications of the distinguishing features are unclear. Although protein interaction networks and connectome exhibit degree assortativity and $m4\_4$ motif, namely $4-$cycle, as the most distinguishing features, their structural variability which is displayed as the spread of points makes it infeasible to even hypothesize the relationship between those features and underlying mechanism of such networks. For metabolic networks, the underrepresentation of $m4\_6$ motif ($4$-path) is found to be the most distinguishing characteristic for the category. However, connecting this result with the possible mechanism which yields the underrepresentation of the motif in metabolic networks is yet under investigation.

 In this analysis we have successfully identified a set of distinguishing structural features for some network sub-domains, such as ecological food webs, online social networks and communication networks and these features seem to coincide with the results from previous studies of such sub-domains. We have also pointed out that the distinguishing features have some implications for the underlying mechanism of networks of sub-domains.

\subsection*{Characterizing the structural diversity of different types of networks.}
Every network belongs to some sort of network domain or sub-domain that usually describes the properties and even structure of a network. As we have seen in the previous section, some networks of selected representative sub-domains exhibit structural uniqueness which makes them stand out among others in the feature space. However we can also observe the overlaps between networks of representative sub-domains and networks of other sub-domains in the Fig.~\ref{2d_figures}, which leads to the question we have asked:
\begin{center}
\textit{Are there any sets of network categories that are inherently indistinguishable from each other based on network structure?}
\end{center}
In this section we explore the structural similarities of different kinds of network domains and sub-domains using machine learning techniques such as random forest and confusion matrix.  

\medskip \paragraph*{Experimental settings.}
We derive structural similarity between network \mbox{(sub-) domains} from a confusion matrix that describes when a random forest classifier makes mistakes and when it does not. However, due to the nature of the classification algorithm and randomly splitting the data into training and testing sets, there involves some randomness in a confusion matrix every time one runs the analysis. Therefore, in order to remove the factor of randomness as much as possible, we run the analysis 1000 times and average the outcomes, namely confusion matrices. Averaging confusion matrices is done element-wise: an element of averaged confusion matrix, say $c$ is defined as $c = \frac{1}{1000}*\sum_{i=1}^{1000} c_i$, where $c_i$ is the corresponding element of an $i$th confusion matrix.

In order to control and compare the impact of class imbalance problem, we use four sampling methods: (i)~no-sampling, namely running the analysis on the original data set; (ii)~random over-sampling in which minority classes are over-sampled to an extent where all classes have the same number of instances as the largest class; (iii)~random under-sampling in which majority classes are under-sampled to an extent where all classes have the same number of instances as the smallest class; (iv)~SMOTE in which all minority classes have synthesized new instances so that the number of data points equals to the one of the largest class. A set of features includes, as before, clustering coefficient, degree assortativity and $k = 4$ connected network motifs which result in eight features.

\begin{figure*}
\centering 
\subfloat[No sampling]{%
  \includegraphics[width=0.89\columnwidth]{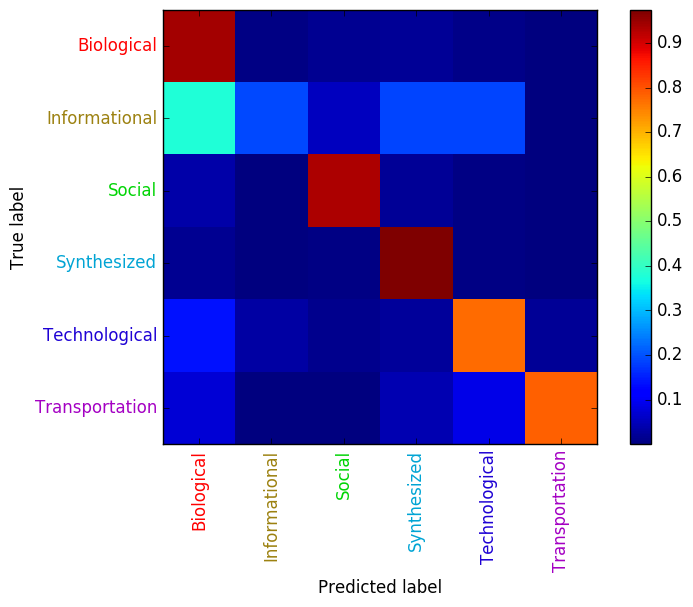}%
   \label{no_confusion}%
}
\subfloat[Random over-sampling]{%
  \includegraphics[width=0.89\columnwidth]{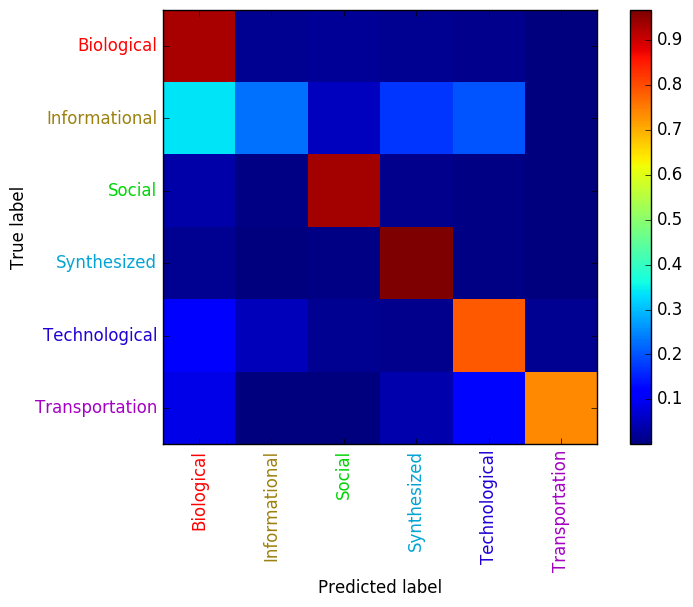}%
  \label{random_over_confusion}%
}

\medskip

\subfloat[Random under-sampling]{%
  \includegraphics[width=0.89\columnwidth]{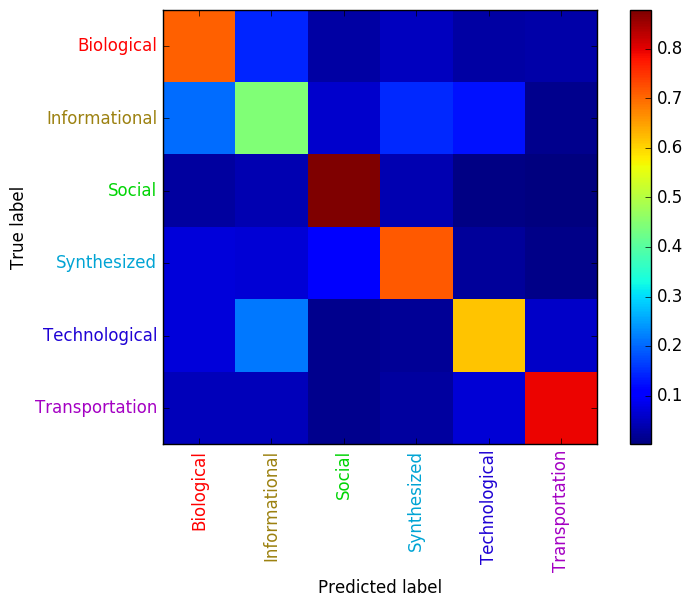}%
   \label{random_under_confusion}%
}
\subfloat[SMOTE]{%
  \includegraphics[width=0.89\columnwidth]{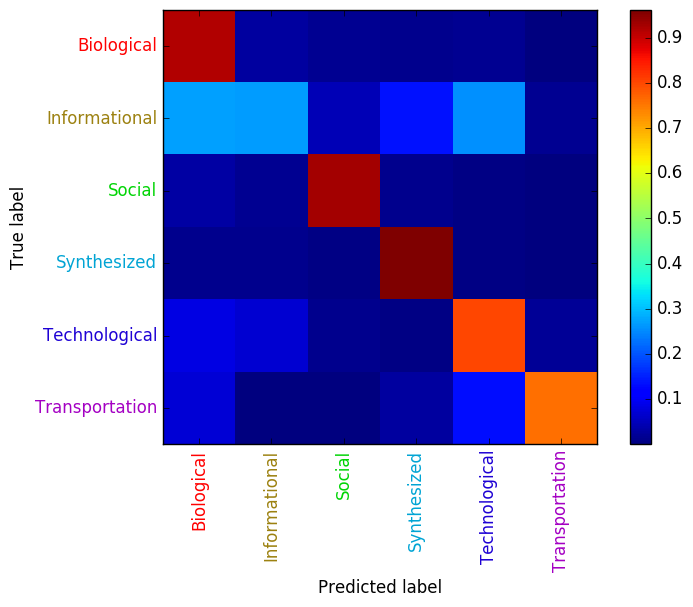}%
  \label{smote_confusion}%
 }
\caption{Confusion matrices for each of the sampling strategy. The color of each cell represents a count value normalized by the sum of all counts in a row to which the cell belongs. Our similarity measurement is based on this normalized value of a confusion matrix. } \label{confusion}.
\end{figure*}

\medskip \paragraph*{Distinguishing networks by domain.} 
We first proceed to work on classification of network domains that include Biological, Social, Informational, Synthetic, Technological and Transportation. Fig.~\ref{confusion} shows the aggregated confusion matrices for each of the sampling strategies. In an aggregated confusion matrix, each cell represents the averaged frequency that an instance of class $i$ is classified as class $j$ in 1000 experiments. As the majority classes inherently contain a number of test data points that leads to a larger count in the confusion matrix, there needs to be some normalization so that each class of network domains becomes comparable with others. The normalization in this study is defined as the following: Let $c_{ij}$ be an element in a confusion matrix. We normalize this quantity by a summation of elements in a row $c_{ij}$ belongs to, namely $\sum_{j=1}^N c_{ij}$, where $N$ is the number of network domains.

The diagonal elements of confusion matrices shown in Fig.~\ref{confusion} indicate the correct classifications. In every sampling method, instances of all network domains are relatively classified correctly, which is observable from the colors of the diagonal cells, except Informational networks for which we observe unsuccessful classifications. In spite of the fact that for random-under sampling there are only 26 instances for the training set of each domain, the confusion matrix for the sampling method still exhibits a strong diagonal pattern, which may imply that instances of various network domains are inherently quite separable in the feature space.

Although it is a meaningful result that the instances of network domains may inherently be separable, our focus now moves on to the off-diagonal elements of confusion matrices. In all matrices, a number of instances of Informational networks are classified as other domains, such as Biological, Synthesized and Technological which is observable from elements of a row corresponding to  ``true Informational.'' Also some instances of network domains are classified as Biological networks, observed in elements of a column corresponding to ``predicted Biological.''

These pieces of information imply the existence of underlying similarities within network domains. However, it is hard to perceive the structure of inter-domain similarity just by looking at the confusion matrix. Therefore, we construct a network of network domains from a similarity matrix (weighted undirected adjacency matrix) that is constructed based on a confusion matrix that is shown in Fig.~\ref{meta_network}. The operation for constructing the similarity matrix based on the normalized confusion matrix is straightforward: we symmetrize the matrix by taking the maximum of two elements in the matrix that are symmetric to each other with respect to a diagonal line. In all cases Biological, Informational and Technological domains are connected with wide edges together, indicating their structural similarities derived from confusion matrices are quite high. From the figure, it appears that there is no ``winning'' sampling method that produces an outstanding result of a domain network, which means that all of the sampling methods practically produce domain networks that are practically the same.

This analysis which is based on network domains is informative in a sense that it consistently exhibits the well connected group of domains that includes Biological, Informational and Technological. However, each network domain includes sub-domains within itself and these network sub-domains are quite diverse in terms of network's function. For example, neural networks in a brain and ecological food web, both in Biological domain, function very differently. Grouping sub-domains of different functions together as a single category may lose some information local to a specific sub-domain. Therefore, we proceed to analyze the networks on a more fine-grained setting, namely using network sub-domains as the class label in classification tasks.


\begin{figure*}
\centering 
\subfloat[No sampling]{%
  \includegraphics[width=0.85\columnwidth]{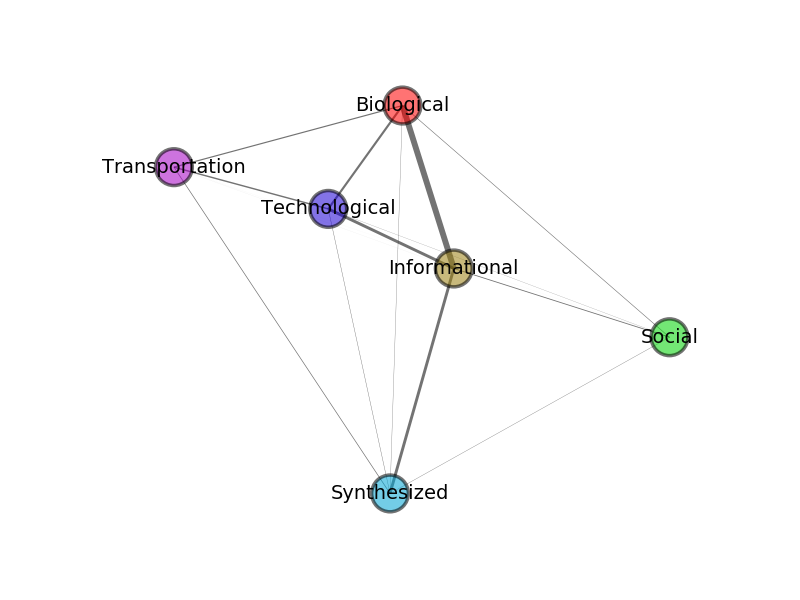}%
   \label{no_graph}%
}
\subfloat[Random over-sampling]{%
  \includegraphics[width=0.85\columnwidth]{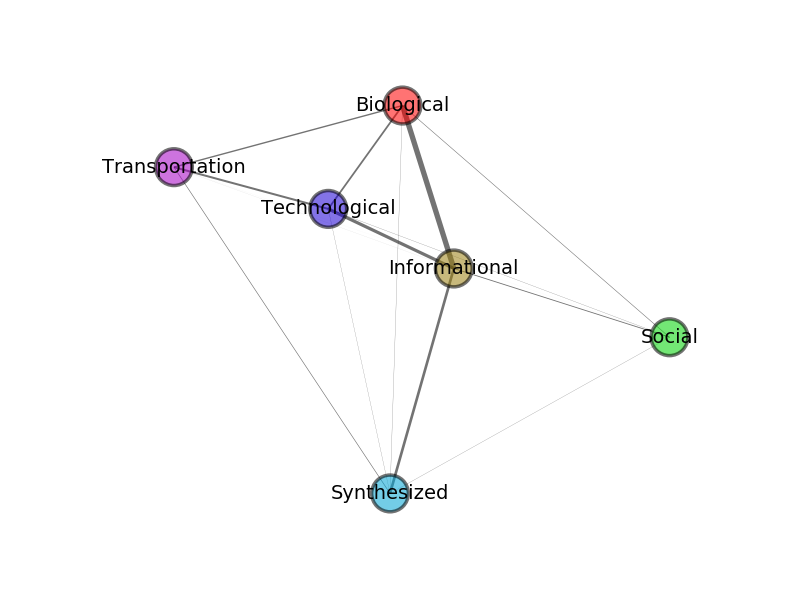}%
  \label{random_over_graph}%
}

\medskip

\subfloat[Random under-sampling]{%
  \includegraphics[width=0.85\columnwidth]{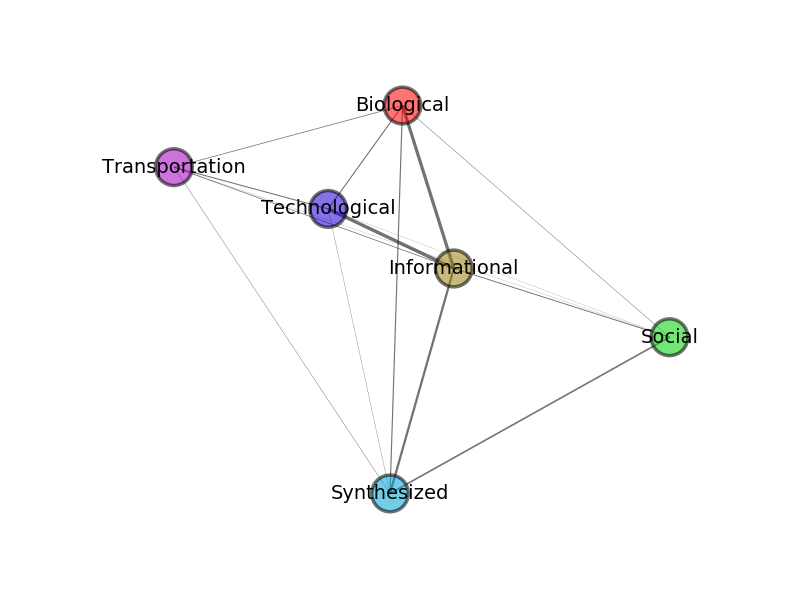}%
   \label{random_under_graph}%
}
\subfloat[SMOTE]{%
  \includegraphics[width=0.85\columnwidth]{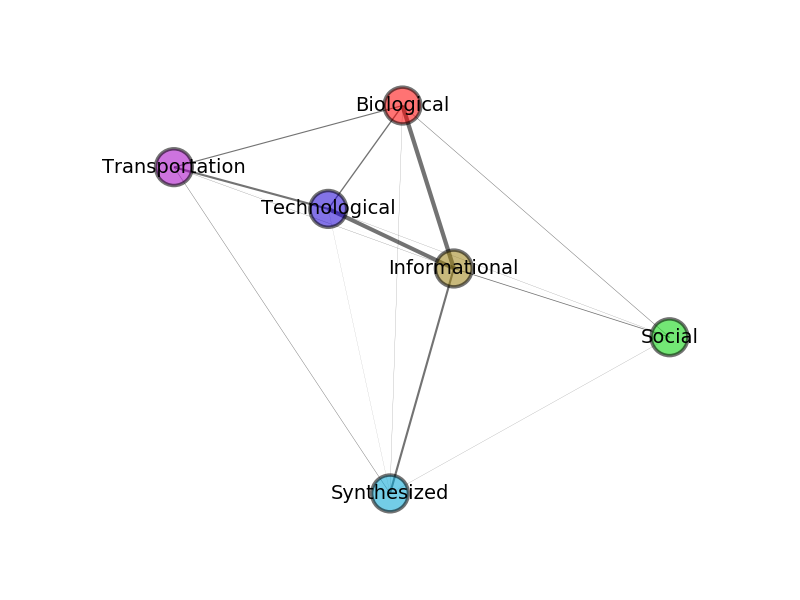}%
  \label{smote_graph}%
 }
\caption{Domain networks for different sampling strategies. The widths of edges are proportional to the magnitude of a value in a symmetrized, normalized confusion matrix. Biological, Informational and Technological domains are well connected in all cases.} \label{meta_network}.
\end{figure*}

\medskip \paragraph*{Distinguishing networks by subdomain.} 
Here we continue the same analysis as we have done in the previous section, but using network sub-domains as class labels for classification tasks.

Although in total we have 33 network sub-domains as shown in Fig.~\ref{sub_dist}, some sub-domains have only few instances, which makes it infeasible to proceed to classification tasks as they involve train-test split of data and SMOTE assumes a training set to have enough amount of instances for selecting $k$ nearest neighbors. Therefore we first exclude sub-domains that the number of instances within themselves are less than seven, which enables us to split the training and test sets in the ratio of $7:3$ while keeping the class distribution in both sets the same and SMOTE to successfully find $k=3$ nearest neighbors for any sub-domain. This filtering results in 22 network sub-domains for this analysis~\footnote{The excluded network subdomains are: Language, Relatedness, Legal, Commerce, Recommendations, Collaboration, Fiction, Relationships, Email, Power grid and Airport.}. 

\begin{figure*}
\centering 
\subfloat[No sampling]{%
  \includegraphics[width=0.9\columnwidth]{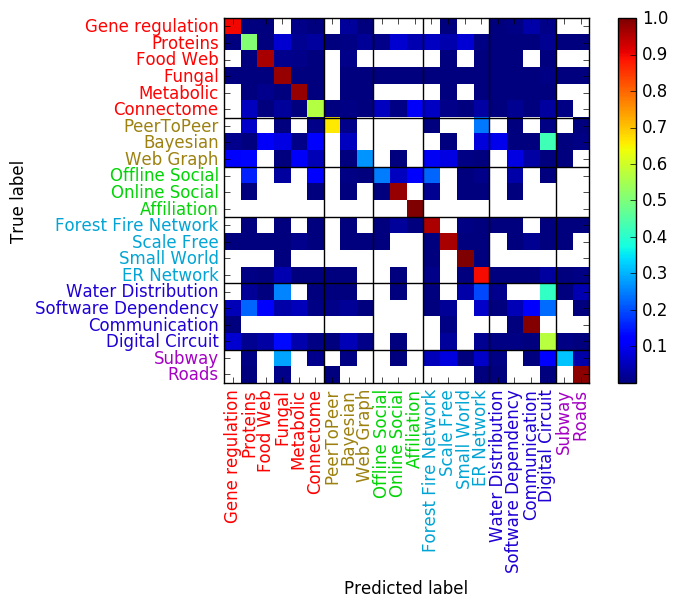}%
   \label{no_confusion_sub}%
}
\subfloat[Random over-sampling]{%
  \includegraphics[width=0.9\columnwidth]{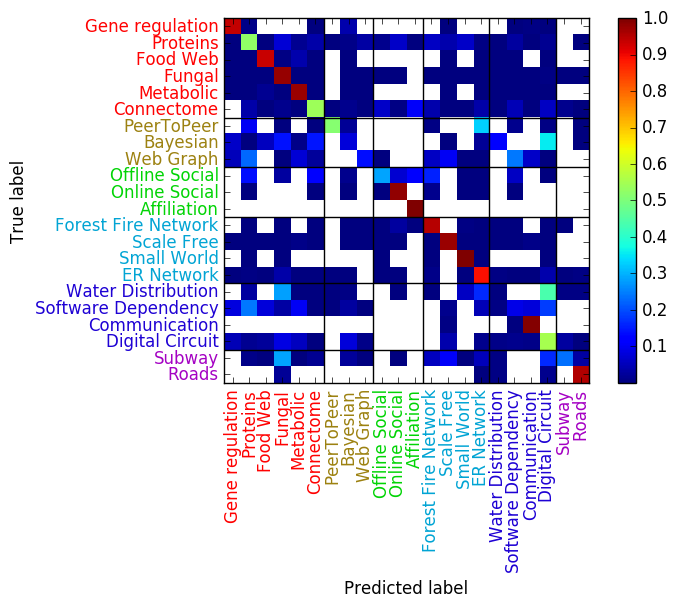}%
  \label{random_over_confusion_sub}%
}

\medskip

\subfloat[Random under-sampling]{%
  \includegraphics[width=0.9\columnwidth]{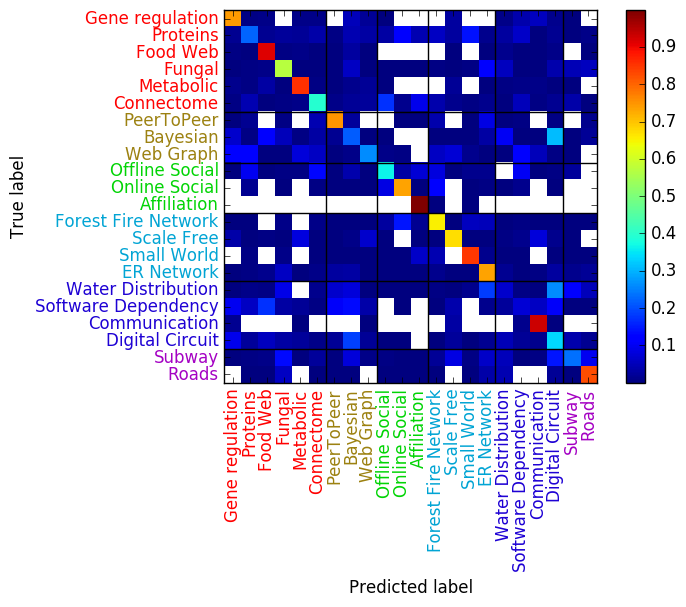}%
   \label{random_under_confusion_sub}%
}
\subfloat[SMOTE]{%
  \includegraphics[width=0.9\columnwidth]{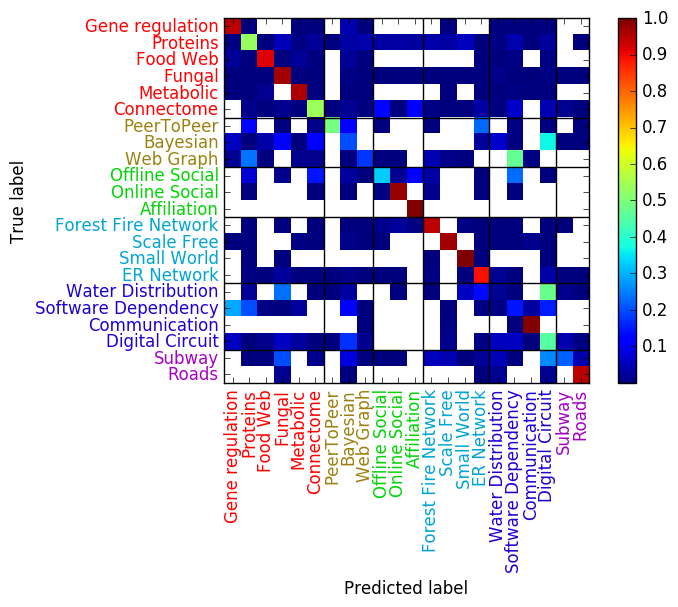}%
  \label{smote_confusion_sub}%
 }
\caption{Confusion matrices for each of the sampling method. The white cells in confusion matrices indicate zero occurrence of corresponding classifications: sub-domain $i$ is misclassified as sub-domain $j$. The color of each cell represents a count value normalized by the sum of all counts in a row the cell belongs to. Our similarity measurement is based on the normalized value of a confusion matrix shown in above. Lines within confusion matrices indicate the separations of network domains.} \label{confusion_sub}.
\end{figure*}

Same as the previous setting for network domain, we run classification tasks 1000 times for each sampling method, aggregate, normalize and symmetrize the resulting confusion matrices. Figure~\ref{confusion_sub} shows the aggregated confusion matrices for each sampling method in which we can observe that all of the confusion matrices exhibit the diagonal pattern, an indication of underlying separability of network sub-domains. Note that, however, some sub-domains such as bayesian, web graph, offline social, water distribution and software dependency, are often not classified correctly which may be due to the few instances and/or the fact that network structures of their instances are actually quite similar to the ones of other sub-domains. It is obvious to notice that the aggregated confusion matrix for random under-sampling displays fewer white cells within itself, meaning that under this sampling method the classifier confused a number of sub-domains with other sub-domains.

In order to visualize the underlying similarities within network sub-domain, we again construct networks of sub-domains based on a weighted undirected adjacency matrix derived from aggregated confusion matrices, shown in Fig.~\ref{meta_network}. From this figure, we can observe that the network domain is not necessarily a good indicator of sub-domain clustering: informational networks including web graph, bayesian and peer to peer network and biological networks, such as metabolic, fungal, food web, etc. are not clustered together, meaning that we do not observe the sub-domains having the same color forming a community together. This could be attributed to the fact that some network domains, such as biological networks, entail a broad spectrum of network sub-domains within itself whose instances are drastically different in terms of the physical size of the things nodes represent (from cells to animals) and the process of networks (from chemical reactions to prey-predator relationships). 

Then the question is:
\begin{center}
\textit{What could be a good indicator of similarity in networks of different kinds?}
\end{center}
In order to answer this question, we first need to discover the communities of subdomain networks which are groups of nodes within which the weighted edge density is high, but between which the weighted edge density is low.

\begin{figure*}
\centering 
\subfloat[No sampling ($4$ communities)]{%
  \includegraphics[width=1.05\columnwidth]{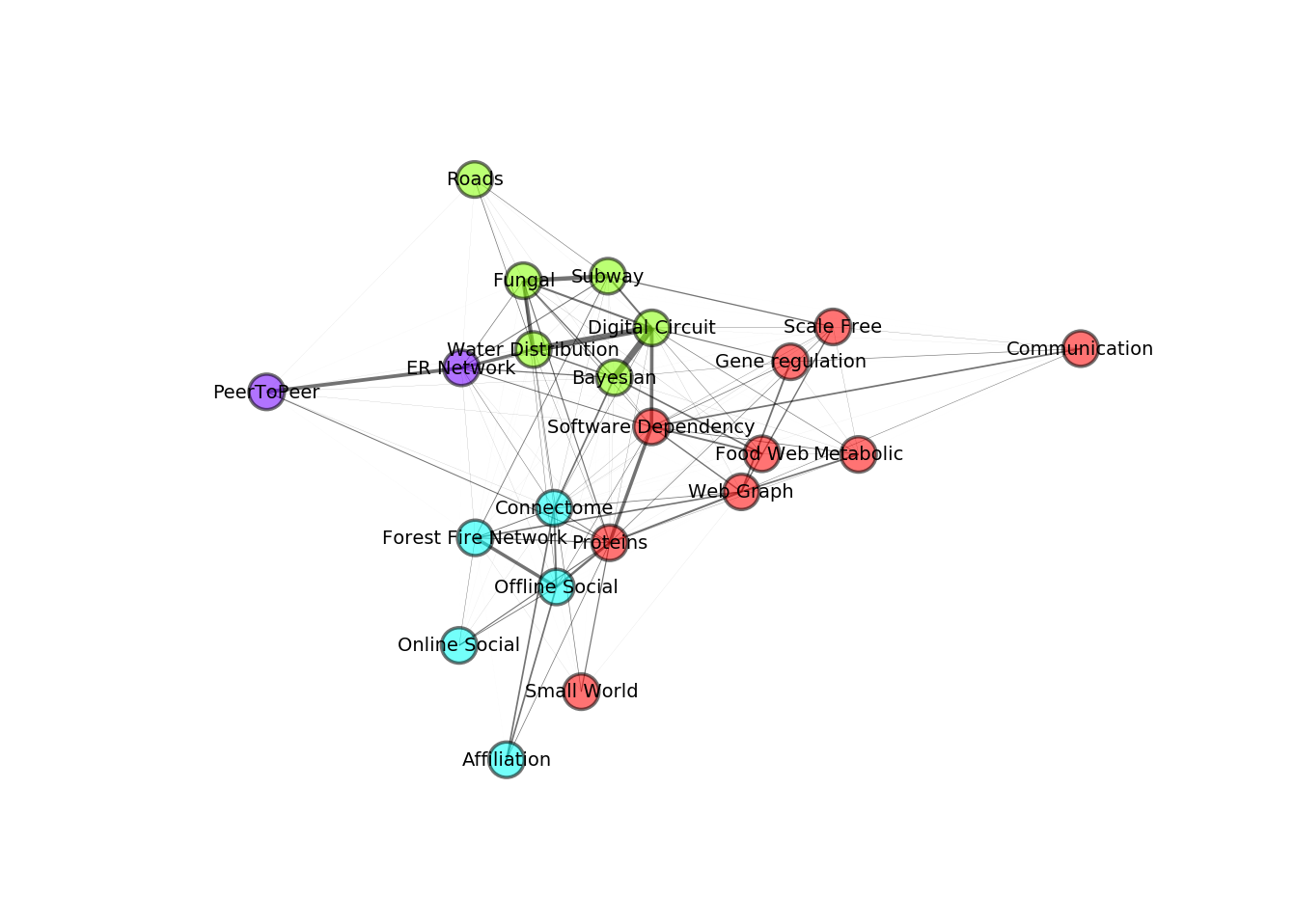}%
   \label{no_graph_sub_original}%
}
\subfloat[Random over-sampling ($4$ communities)]{%
  \includegraphics[width=1.05\columnwidth]{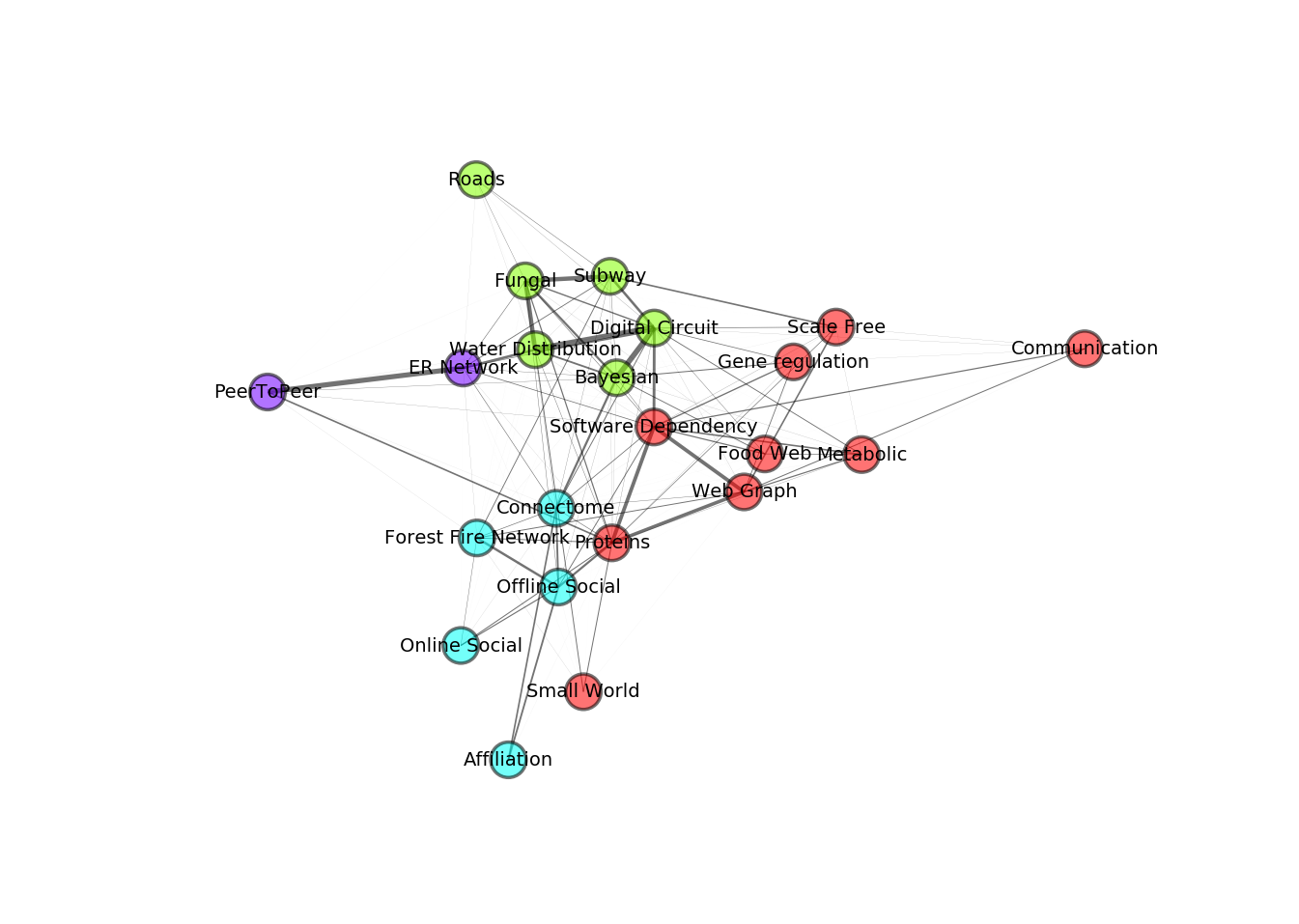}%
  \label{random_over_graph_sub_original}%
}


\vspace{-4mm}
\subfloat[Random under-sampling ($3$ communities)]{%
  \includegraphics[width=1.05\columnwidth]{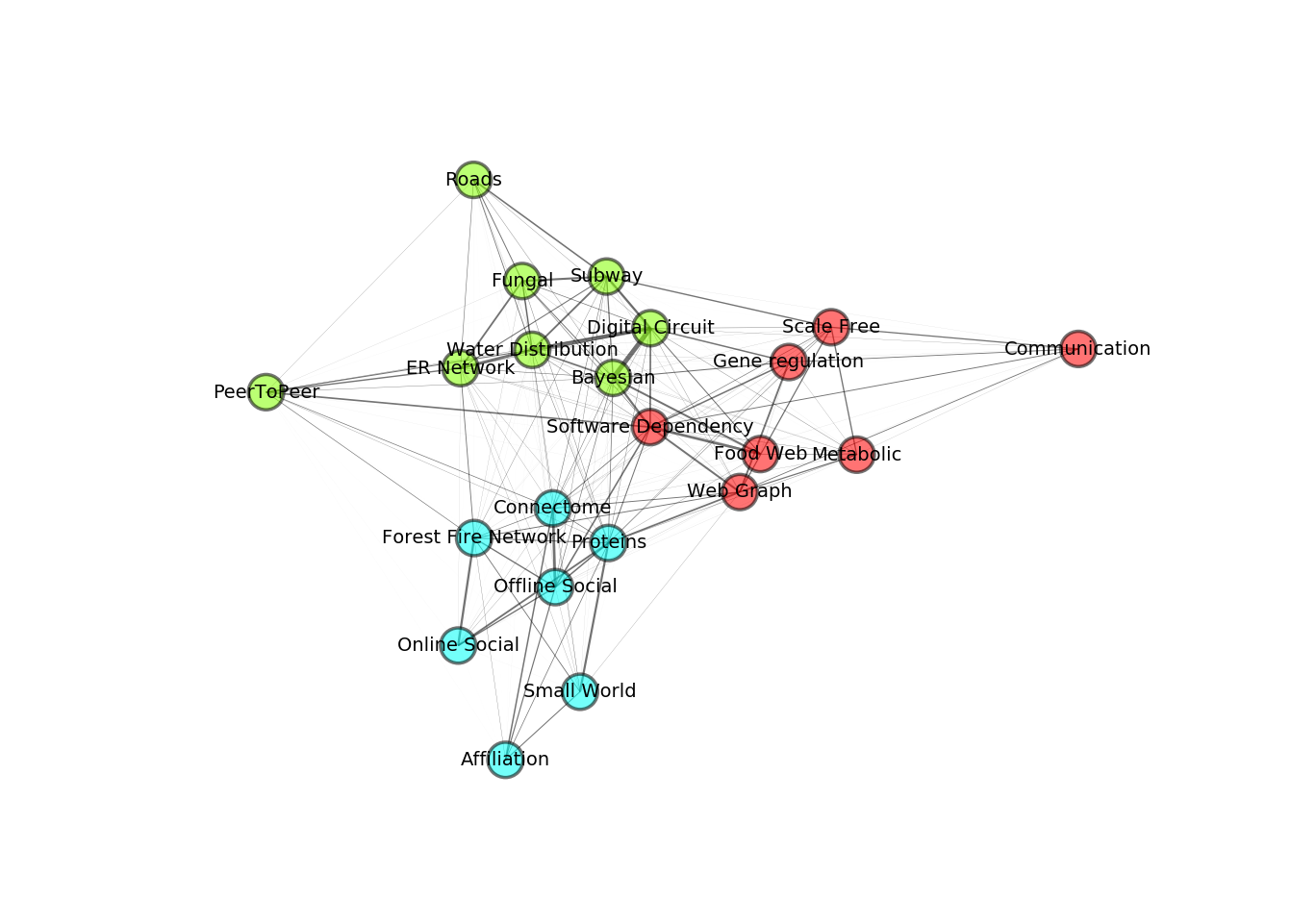}%
   \label{random_under_graph_sub_original}%
}
\subfloat[SMOTE ($4$ communities)]{%
  \includegraphics[width=1.05\columnwidth]{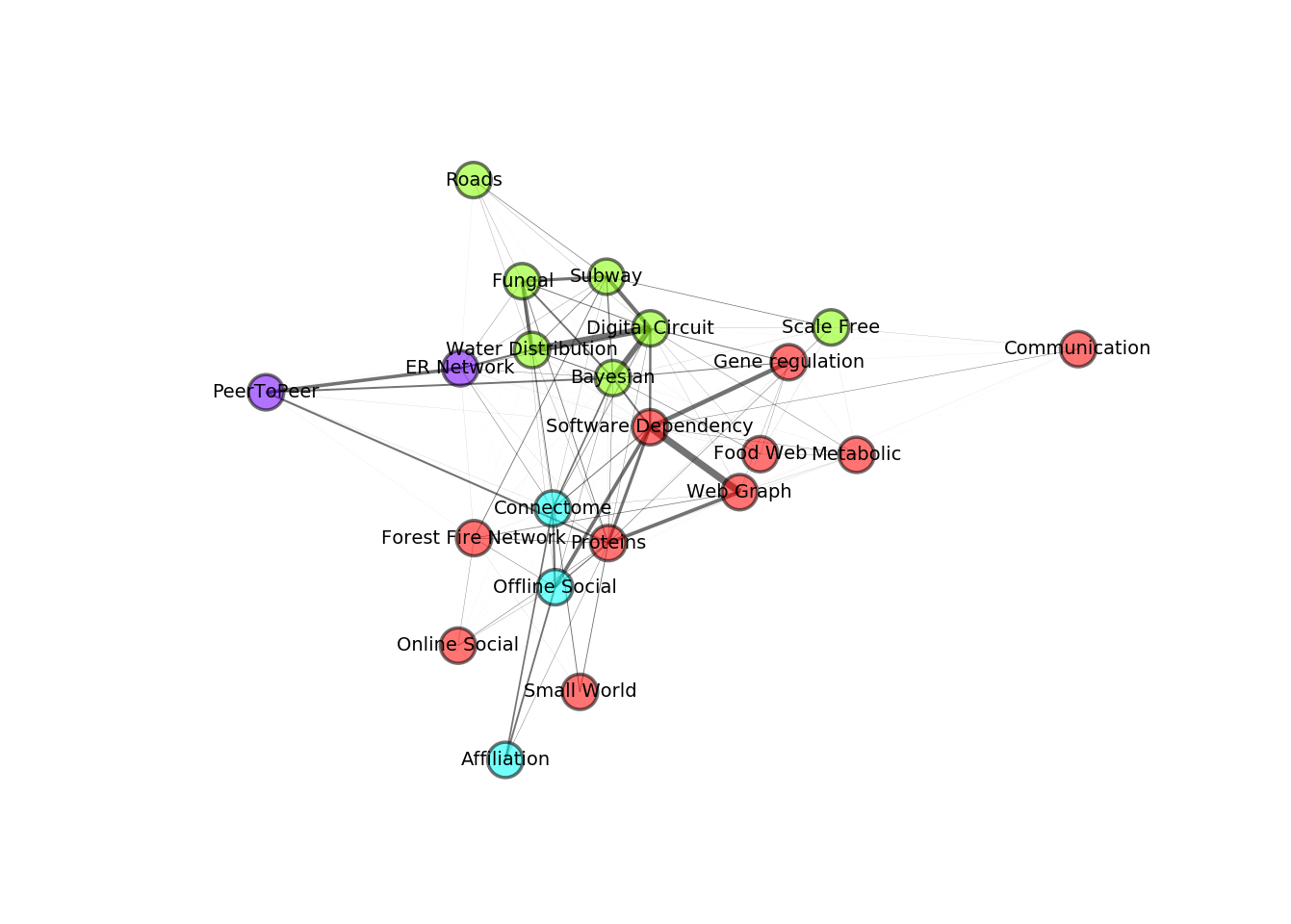}%
  \label{smote_graph_sub_original}%
 }
\caption{The networks of sub-domains with community labelings. The color of a node now corresponds to a community found by the algorithm and width of each edge corresponds to the similarity of sub-domains derived from the aggregated confusion matrix.} \label{meta_network_community}.
\end{figure*}

Fig.~\ref{meta_network_community} shows the networks of sub-domains on which the colors of nodes correspond to the community membership found by a community detection algorithm proposed by Clauset \textit{et al.} \cite{CNMAlgorithm}. We have used an implementation of the modified version of this algorithm for weighted network that is available in Python-igraph as a method \texttt{community\_fastgreedy()} \cite{igraph}. From Fig.~\ref{meta_network_community}, one may notice that the community structure across different sampling methods is almost consistent, meaning that the certain groups of sub-domains are always in the same community.  For instance, for all sampling methods, offline social, connectome and affiliation networks are in the same community. 

Some network sub-domains, however, change the community membership for some subdomain networks. For instance, online social network and forest fire model networks join the community of off-line social network for no sampling, random-over sampling and random-under sampling method, but joins the community of software dependency (red color) for SMOTE sampling. This variation of community membership across all subdomain networks could be attributed to the effect of each sampling method on decision surfaces a classifier builds for classification. For instance, decision surfaces built by a classifier under the random under-sampling strategy should be simple shape-wise since the training set is very sparse in terms of the number of data points. On the other hand, decision surfaces built by a classifier for random over-sampling should be more toward complicated shape-wise and lead to over-fitting as the number of data points for training is larger and duplicated points force the classifier to adjust itself to those points. Therefore this variation of decision surfaces may lead some data points to be classified differently for different sampling method, which ultimately leads to the variation of community memberships.

\begin{figure*}
\centering 
\subfloat[No sampling]{%
  \includegraphics[width=1.05\columnwidth]{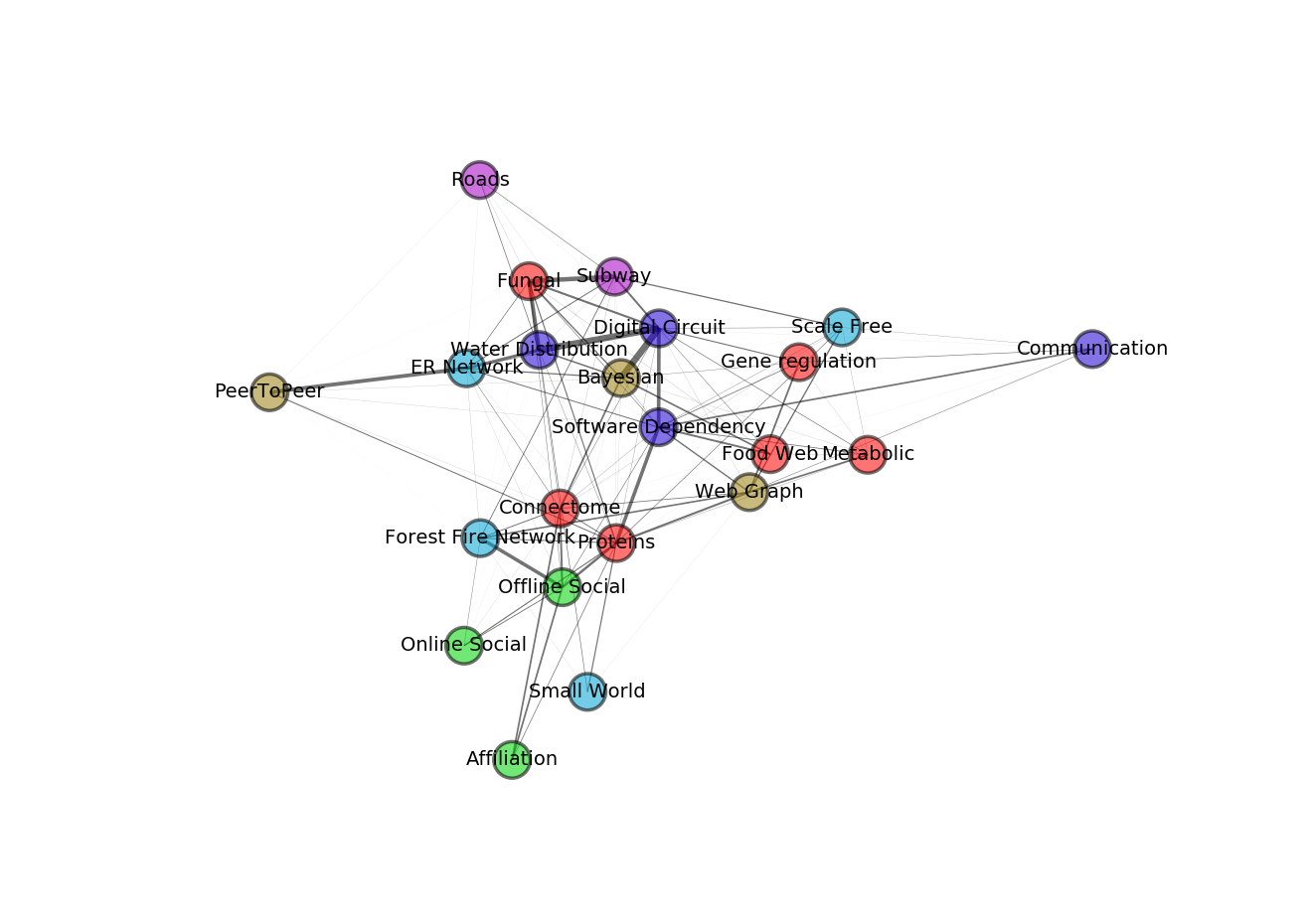}%
    \label{no_graph_sub_original}%
}
\subfloat[Random over-sampling]{%
  \includegraphics[width=1.05\columnwidth]{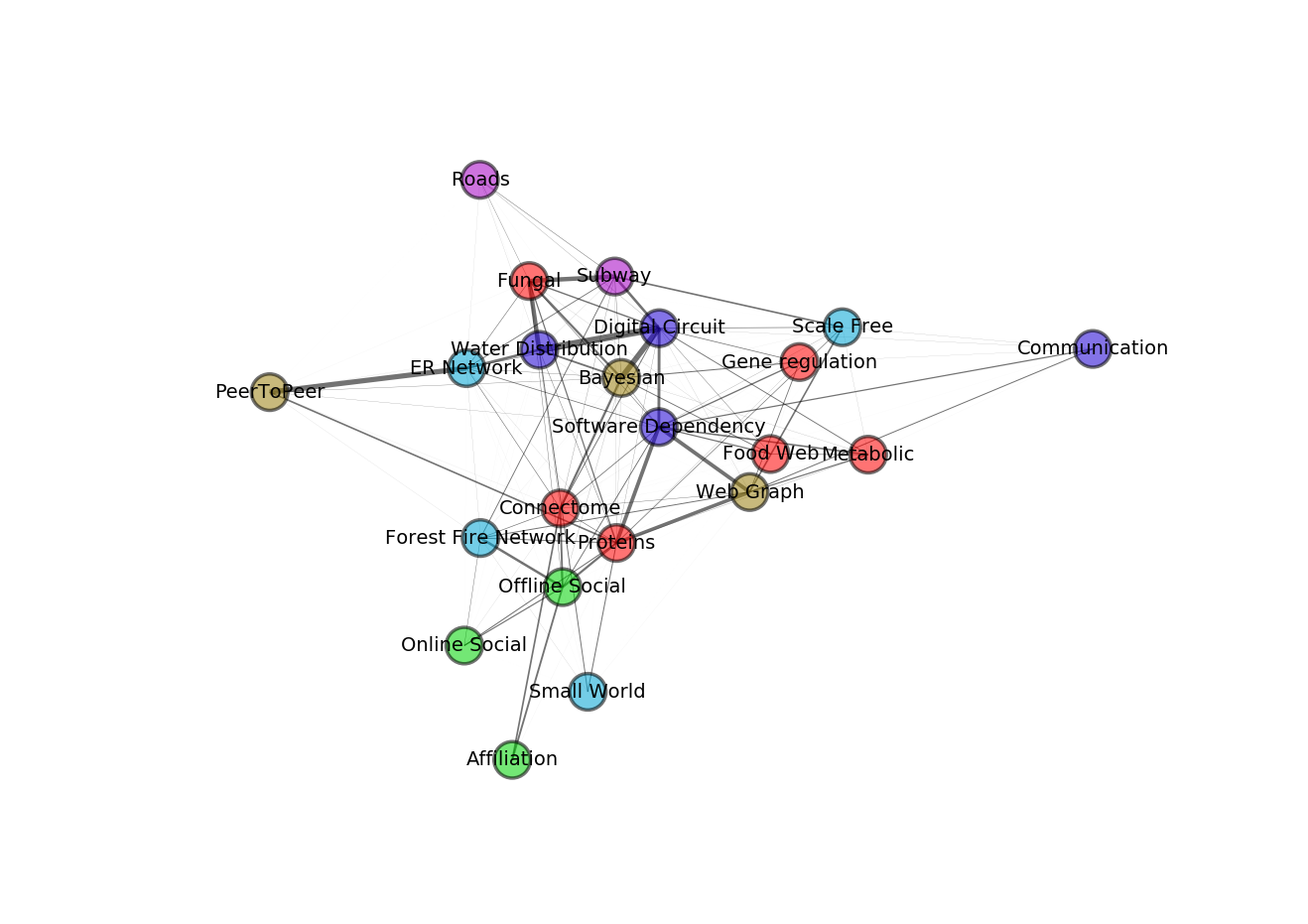}%
   \label{random_over_graph_sub_original}%
}


\vspace{-4mm}
\subfloat[Random under-sampling]{%
  \includegraphics[width=1.05\columnwidth]{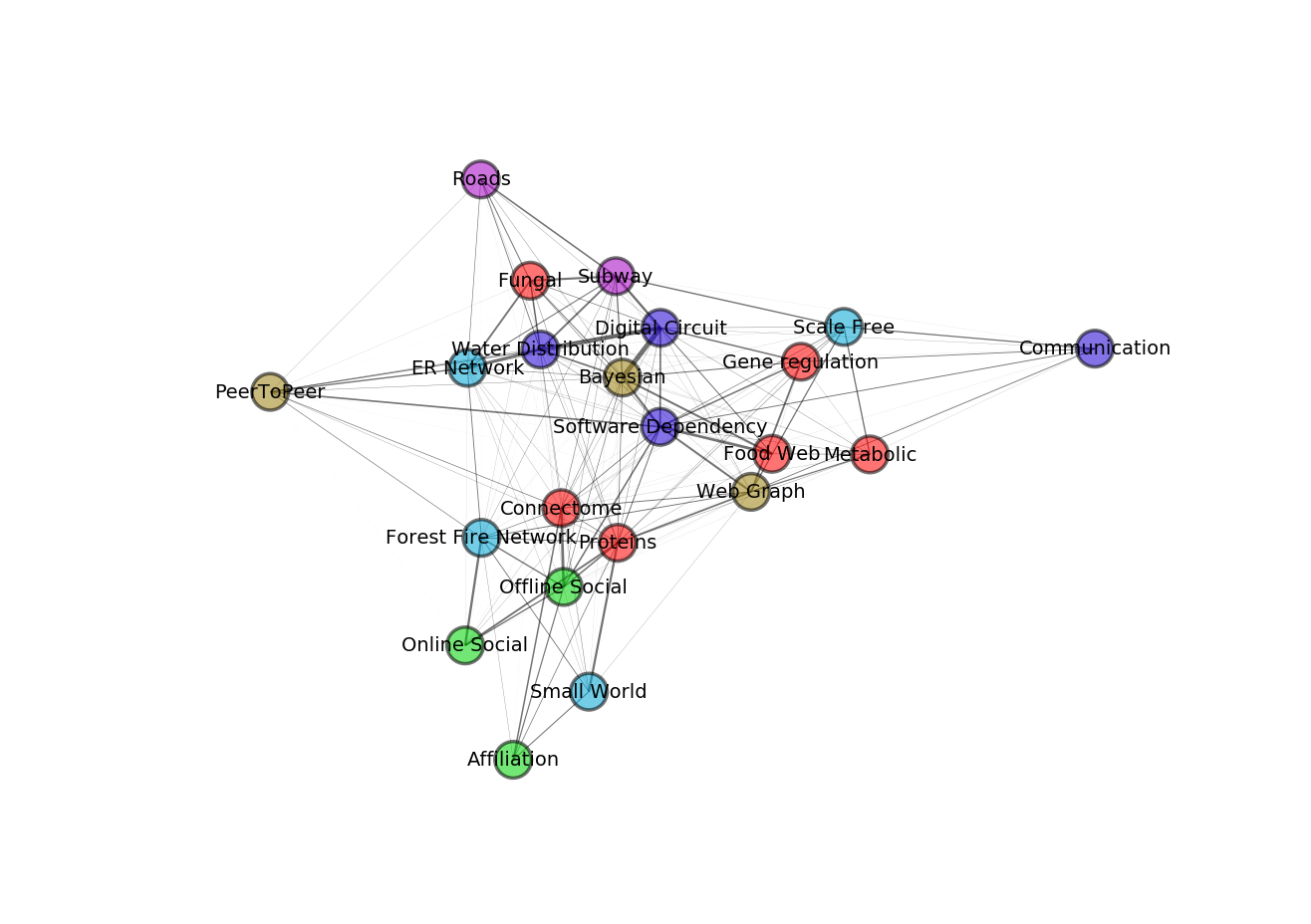}%
   \label{random_under_graph_sub_original}%
}
\subfloat[SMOTE]{%
  \includegraphics[width=1.05\columnwidth]{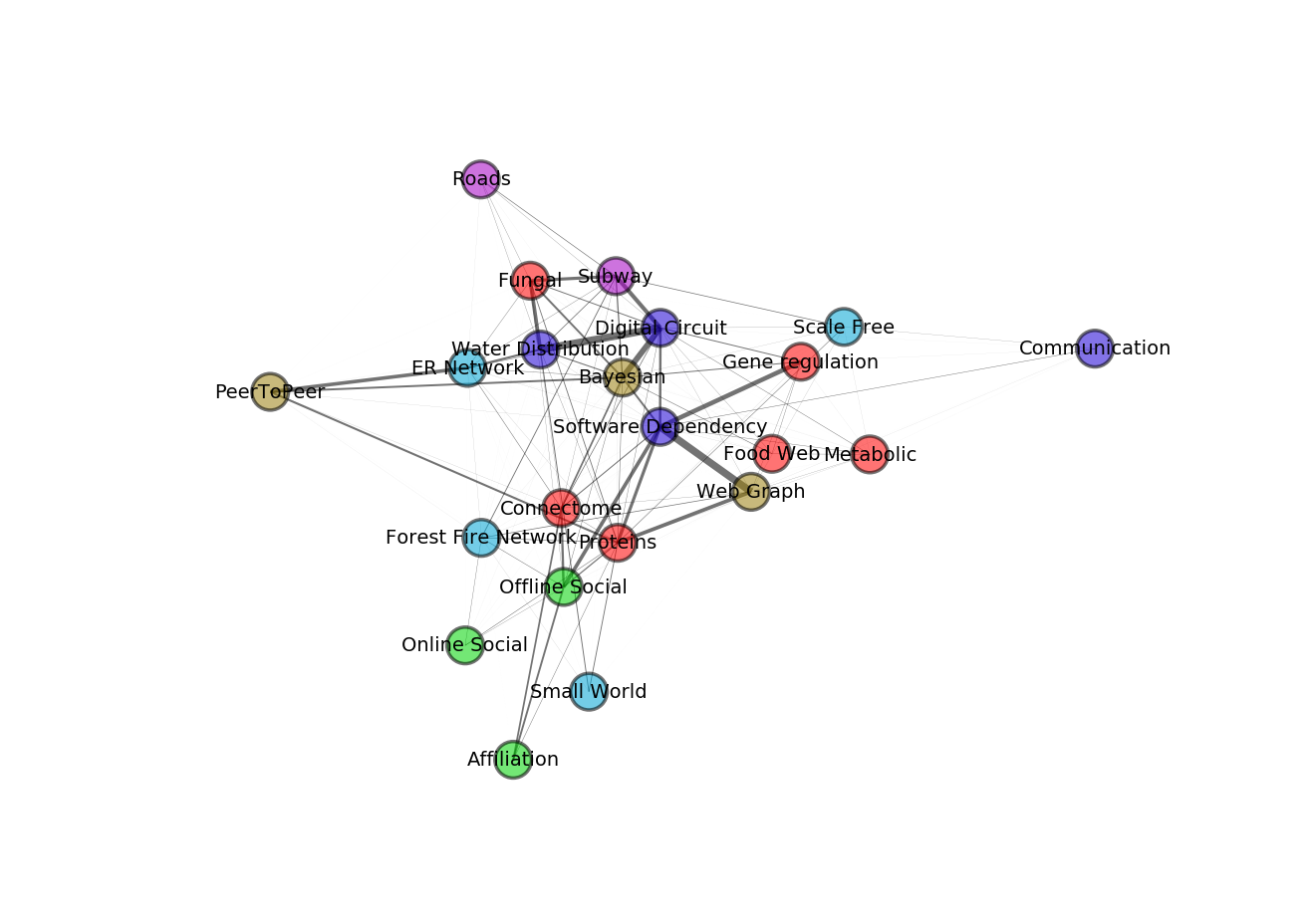}%
   \label{smote_graph_sub_original}%
 }
\caption{The networks of sub-domains. The color of each node corresponds to a domain the sub-domain (node) belongs to and width of each edge corresponds to the similarity of sub-domains derived from the aggregated confusion matrix.} \label{meta_network}.
\end{figure*}


\begin{figure}[ht]
		\begin{center}
		\includegraphics[clip,width=9cm,height = 6cm]{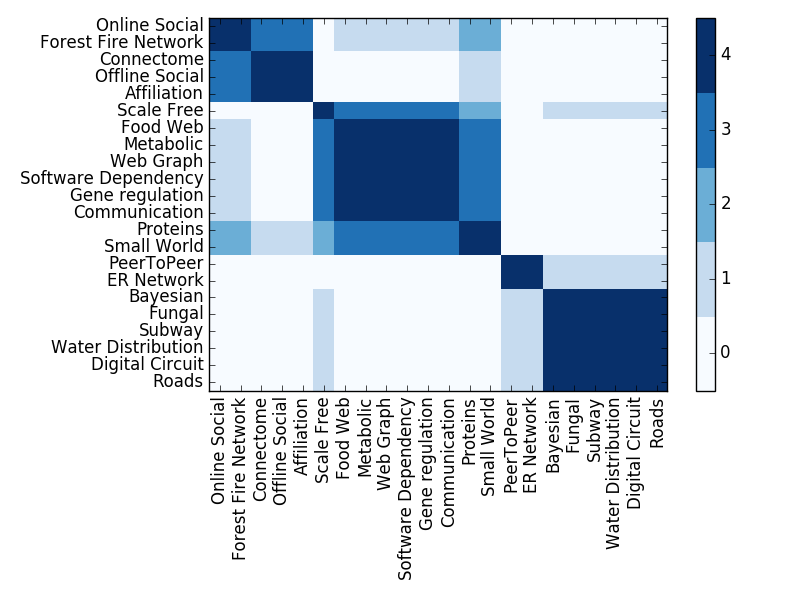}
		\vspace{-4mm}
		\caption{Overlaps of community memberships. The intensity of color for each cell in the matrix indicates the frequency of two sub-domains being in the same community.  There are three groups of sub-domains that almost constantly share the community membership: (i) community of social networks including online social, forest fire model, offline social, and so on; (ii) community of networks claimed to have a power-law degree distribution including scale-free network, metabolic network, web graph, etc. with some exceptions such as small world network; (iii) community of networks that are physically embedded, having an input-output function or containing some sort of flow on the network, which include road, digital circuit and fungal networks.}
		\label{community_overlaps}
 		\end{center}
\end{figure}

In order to facilitate analyzing the variation of community memberships, we have constructed a matrix, shown in Fig.~\ref{community_overlaps} in which the frequency that two sub-domain share the same community membership corresponds to the color intensity. For instance offline social, connectome and affiliation networks are assigned to the same community four times. From this figure, one may notice that there appears to be three groups of sub-domains that are consistently assigned to the same community. The first one is a community of social networks including online and offline social networks as well as affiliation network and forest fire model with an exception of connectome, namely brain networks. This grouping of social networks could be attributed to the same underlying process, namely forest fire process: on online social networks, namely Facebook networks, one becomes a friend with someone, and finds other friends on the person's friend list, sends them friend requests and recursively continues the process; on off-line social networks, a person introduces his/her friends to you, they also introduce their friends to you and the process recursively continues. The possible explanations why we observe connectome being in the same community as such social networks are as follows: (i) the lack of feature or dimension in the feature space that distinguishes brain networks from social networks; (ii) the underlying generative mechanism of brain network is actually similar to that of social networks. As far as we know, however, there is no previous study investigating commonality of processes of both social and brain networks. Therefore it is reasonable to speculate here that we lacked a set of distinguishing features for connectome, which resulted in the clustering of the network with social networks.

The second group of sub-domains corresponds to the networks that have been claimed to have the power-law degree distribution which include scale-free network, metabolic network, web graph, etc. and some of these networks have been conjectured to grow according to a mechanism called preferential attachment. In this network generative process, newly added nodes in a network, for example newly created web page or software package, tend to connect to the popular or high-degree nodes, i.e. popular web sites or widely used software packages. The third and last group of sub-domains is the network of ``flow'' that includes electrical signal (digital circuit), information (bayesian), water (water distribution), nutrient (fungal), people/trains (subway) and cars (road).  Also, if we look at the subdomain networks, digital circuit and bayesian networks are always tightly connected together. This may be due to the fact that they are both an input-output network as well as being a flow network. The rest of the flow networks have another common property, that is, physical embedding of the network. These networks have a strong constraint of the physical limitation. For instance, it is almost impossible for a node in a physically embedded network to have a thousand connections upon it.

From these communities of sub-domains, one could hypothesize an idea that explains what governs the structure of complex networks: functionality, constraint and growing mechanism of networks may play an important role of the formation of a specific network structure. As we have seen, the sub-domains of networks that have the similar functionality, constraint or growing mechanism exhibit the similar structural pattern which is captured in confusion matrices and also communities of networks. However, there seems to be a case where we do not know why a network sub-domain is in a community and how the hypothesis would explain this: connectome, or brain network shares the same community with social networks. If our hypothesis is correct, then brain networks should have either the same function, constraint or growing mechanism as social networks. As far as we know, there is no study which compares connectome and social networks in terms of network structure and explains the similarity of their function, constraint and generative mechanism. One possible direction to take would be finding a set of features for which those sub-domains have similar values and investigate a possible generative process which yields networks having such structural features.

Although there are some exceptional cases, our findings from this experiment provide some supporting evidence that the network structure may be influenced by the underlying function, constraint and growing mechanism of the network.

\section*{Discussion}
In the previous section, we have found the distinguishing features for various network sub-domains with possible explanations for the underlying processes of networks, and the hidden similarities among network domains and sub-domains based on subdomain networks we have constructed. Here, we synthesize our findings and the previous studies together.

In the investigation of distinguishing features and the separability of sub-domains, we have observed that some network sub-domains are hard to be separated in the high dimensional feature space, which can be seen in the AUC score. This finding has an interesting connection with the subdomain networks we have constructed later: the sub-domains having high separability, such as online social networks and ecological food-webs, tend to be at the \textit{periphery} of the subdomain networks, whereas the subdomains having low separability, such as protein interaction networks and connectome, tend to be at the \textit{core} of the subdomain networks. In the subdomain networks, the core of nodes essentially depicts the sub-domains that are frequently misclassified by a classifier due to their structural similarity and the periphery displays sub-domains that are dissimilar to other sub-domains in terms of network structure. Therefore, it is relatively reasonable to infer that the sub-domains at the core that were not studied for finding the distinguishing features, such as bayesian networks and web graphs, may also exhibit the low separability in the feature space.

 As we have seen in confusion matrices in Fig.~\ref{confusion}, network domains such as Biological, Social, etc. are quite separable in the feature space. Also, confusion matrices of network sub-domains exhibit the strong diagonal patterns except for a few such as bayesian network and web graphs, indicating the separability of networks at the sub-domain level. The separability of networks in the different levels (domain and sub-domain) tells us something about where networks of different types occur within a manifold in the high dimensional feature space we have constructed: in the feature space at the domain level, points that correspond to individual networks in the same network domain occupy some space in an intricately shaped manifold; diagonal elements in the confusion matrices for network domain imply that instances of different network domains occupy different locations within the manifold with some overlap, which could be observed in off-diagonal elements in the confusion matrices. At the network sub-domain level, we see a somewhat different outcome. Some sub-domains in a network domain occupy regions in the feature space that are completely separated. They correspond to the non-overlapping sections of the manifold at the level of network domain. They include, for example, food webs for Biological, peer to peer for Informational, communication for Technological and roads for Transportation. Some network sub-domains, however, almost completely overlap in the manifold with other sub-domains and they correspond to the overlapping sections in the manifold at the network domain level. They include bayesian and web graphs for Informational and water distribution and software dependency for Technological.

This idea of thinking networks as data points in a manifold of complex shape within some feature space has been explored previously. Corominas-Murtra \textit{et al.} have studied the idea of \textit{Hierarchy} in which each axis corresponds to some feature related to the structure of networks, such as tree-ness, feed-forwardness and order-ability \cite{Hierarchy}. In this study, it is shown that different kinds of networks, such as technological, language and neural networks occupy some regions in a feature space. One may notice in this study that some regions in the feature space are not occupied at all by any networks. This observation yields a question about the feature space:
\begin{center}
\textit{Are some regions of a feature space theoretically possible for networks to occupy?}
\end{center}
This question may be answered with the study done by Ugander \textit{et al.} \cite{Ugander:2013}. They have studied a feature space in which each axis corresponds to the subgraph frequency of online social networks and mathematically proved that there are some regions in the feature space that are mathematically infeasible to be occupied. In other words, it is theoretically impossible for networks to have a structural property which corresponds to the region in the feature space. Interestingly, they observed that the real world networks, in this case Facebook networks, only occupy some sections of the theoretically feasible region. Taken together, these studies suggest that networks in the high dimensional feature occupy some regions of the entire possible space that is theoretically feasible. This phenomenon may be due to the fact that the space which is not occupied, yet theoretically feasible region, corresponds to an inefficient structure of a network. Many of the biological networks and technological networks are optimized for a functioning by either natural selection over the course of evolution or effort of designing by a number of engineers and they may push the networks into a certain region in the feature space. The convergence toward a certain region in the feature space seems to happen in both kinds of networks, namely biological and technological networks. This is supported by one of our findings that fungal networks, a kind of biological network developed by a biological process, and water distribution networks, a kind of network designed by engineers, are found to be structurally similar based on the results of confusion matrices. This finding may be due to the fact that their optimization is essentially for efficient flow on the network and cost-reduction of wiring in the networks.

 \section*{Conclusion}
In this paper, we have studied 986 real-world networks along with 575 synthesized networks in order to formulate a hypothesis about the structural diversity of complex networks across various domains and sub-domains.
 
  Our study successfully identified the distinguishing features for some network sub-domains including, metabolic, ecological food web, online social and communication networks and found out some of these features could naturally explain the process in the network of interest. There are sub-domains such as protein interactions and connectomes, however, that seem to be indistinguishable from others with a set of features we have utilized 
  
   Using machine learning techniques such as random forest classifier, confusion matrix and network community detection algorithm, we have found that there are some categories of networks that are hard to be distinguished from each other by a classifier based solely on their structural features and these groups of structurally similar yet categorically different networks in fact seem to have some common properties, such as the same functionality, physical constraints and generative process of the networks. There are, however, some categories of networks that are found to be structurally similar, and yet our hypothesis seems to lack some theoretical basis for explaining the observed phenomenon.
 
 Nevertheless, our study sheds light on the direction to which we could uncover the underlying principles of network structure: the functionality, constraint and growing mechanism of network may play an important role for the construction of networks having certain structural properties.

 There is still some room for our study to be improved. The class imbalance problem, even though we have utilized sampling methods in order to alleviate the problem, is one of the main remaining concerns. We could possibly discover other hidden properties and relationships if more instances were added for sub-domains that are excluded from the analyses due to the lack of instances needed for classification tasks, such as language network, collaboration network, power grid network, etc. Another direction for future research is incorporating other scale-invariant structural features. In this study we have only used a set of eight features. It is possible, however, that adding other dimensions in the feature space may reveal other hidden properties that were not captured in our feature set. 

Lastly, our study could extend a work by Middendorf \textit{et al.} in which they trained a machine learning classifier on instances of various network generative models, classified protein interaction networks of the \textit{Drosophila melanogaster} using, a species of fly, and identified a generative mechanism that was most likely to produce the protein interaction networks \cite{MechanismInference}. Using a broad spectrum of networks of different categories, our study suggests a way to construct and validate a hypothesis regarding which network sub-domains have the common generative process. For example, one could train a classifier on only networks that are constructed by some network generative models, feed the classifier instances of real-world networks and observe which categories of networks are classified as which generative processes.


%

\newpage

	\begin{longtable*}{ l | l | p{9cm} }
	\caption{Network sub-domains and their descriptions.} \label{tab:subdomain}\\
	

 	Sub-domain & Domain& Description \\ \hline 
	 \endfirsthead
	 \multicolumn{3}{l}{\small\it Continued}\\ \hline
	 Sub-domain & Domain& Description \\ \hline \hline
 	\endhead
      Fungal &  Biological & Network of mycelial growth patterns of fungus or slime mold. Nodes are located at hyphal tips, branch points, and anastomoses. Edges represent cords.\\  
      Metabolic &  Biological & Network of chemical reactions of metabolism in a cell.\\  
      Proteins &  Biological & Physical contacts of proteins in a cell or in a living organism.\\  
      Connectome &  Biological & Network of neural connection in the brain at either the level of neuron or the level of anatomical region.\\  
      Food Web &  Biological & The relationships of predator-prey in terrestrial or aquatic animal kingdoms.\\  
      Gene regulation &  Biological & The interactions of molecular regulators that govern the gene expression.\\ 
      Bayesian & Informational & Probabilistic model that contains a directed-acyclic-graph(DAG) in which nodes represent random variables and edges represent conditional dependence among these random variables.\\  
      PeerToPeer &  Informational & A kind of computer networks in which peers (computers) are equally privileged in the network for sharing files.\\ 
      Web Graph &  Informational & Network of World-Wide-Web. Nodes represent web pages and edges represent hyper-links among web sites. \\  
      Language &  Informational & Networks of word adjacency and word association that are extracted from books.\\  
      Relatedness &  Informational & Networks of relatedness, such as similarities among book purchased on online retailers. \\ 
      Legal &  Informational & Network of legal citations. Nodes represent majority opinions written by the Supreme Court of the United States and edges represent citation.\\ 
      Commerce &  Informational & Network of co-purchasing items on websites.\\  
      Recommendations &  Informational & Network of books, where edges represent the frequency that a pair of nodes (books) is co-purchased together.\\  
      Affiliation &  Social & Network of cooperate boards and the directors that sit on them. Networks in this sub-domain are one-mode projected onto individuals.\\  
      Online Social &  Social & Network of friendship online, such as Facebook.\\  
      Offline Social &  Social & Network of friendship or some sort of inter-personal relationships offline.\\  
      Collaboration &  Social & Network of collaborations among people. This sub-domain includes collaboration of scientific papers, music, etc.\\  
      Fiction &  Social & Co-appearance network of fictional characters from books.\\  
      Relationships &  Social & Network of social relationship among managers from tech companies, Florentine families during the Italian Renaissance, etc.\\ 
      Email &  Social & Network of emails.\\ 
      Small World &  Synthesized & Networks generated by Watt-Strogatz model.\\  
      Forest Fire Network &  Synthesized & Networks generated by a Forest Fire Model proposed by Leskovec \textit{et al.} \\  
      ER Network &  Synthesized & Erd\H{o}s-R\'enyi random network. \\  
      Scale Free &  Synthesized & Networks generated by Barab\'asi-Albert model.\\  
      Communication &  Technological & Network of autonomous systems (the Internet).\\  
      Digital Circuit &  Technological & Networks of logical gates connected by wirings.\\ 
      Software Dependency &  Technological & Networks in which nodes represent either class, function or package and edges correspond to dependencies.\\  
      Water Distribution &  Technological & Network of piping and junctions for water distribution system.\\ 
      Power Grid &  Technological & Network of power grid in which nodes correspond to transforms or power relay points and edges represent power lines.\\ 
      Roads &  Transportation & Network of roads where nodes are intersections and edges are roads.\\ 
      Subway &  Transportation & Subway networks of major cities around the world.\\ 
      Airport &  Transportation & Network of airports that are connected by flights between the airports.
      
\end{longtable*}

\end{document}